\documentclass[apj,onecolumn,numberedappendix]{openjournal}

\newcommand{\orcidauthor}[3]{\author{\href{http://orcid.org/#1}{#2$^{#3}$}}}
\usepackage[plainpages=false, colorlinks=true, anchorcolor=blue,
            linkcolor=blue, citecolor=blue, bookmarks=false]{hyperref}
\hypersetup{
    colorlinks=true,
    linkcolor=blue,
    citecolor=blue,
    urlcolor=blue,
    anchorcolor=blue
}

\usepackage[T1]{fontenc}
\usepackage{graphicx}
\usepackage{dcolumn}
\usepackage{bm}
\usepackage{longtable}
\usepackage{adjustbox}
\newcommand{\rthis}[1]{\textcolor{black}{#1}}
\usepackage{float}
\usepackage[caption=false]{subfig}
\usepackage{appendix}
\begin{document}

\title{Search for Gamma-ray emission  from Abell 119 galaxy cluster  using INTEGRAL/ISGRI, COMPTEL, and DAMPE data.}
\email{* ph22resch11006@iith.ac.in}
\email{** shntn05@gmail.com}

\orcidauthor{0009-0007-1942-8794}{Siddhant Manna}{1 *}
\orcidauthor{0000-0002-0466-3288}{Shantanu Desai}{1 **} 
\affiliation{$^{1}$Department of Physics, IIT Hyderabad Kandi, Telangana 502284, India.}





\begin{abstract}
We present a comprehensive search for non-thermal high-energy $\gamma$-ray emission from the nearby merging galaxy cluster Abell~119 ($z=0.044$) using archival observations spanning over seven decades in photon energy. Our analysis combines hard X-ray data from INTEGRAL/ISGRI (30--100~keV), MeV $\gamma$-ray observations from COMPTEL (0.75--30~MeV), and GeV--TeV data from the DArk Matter Particle Explorer (DAMPE; 3~GeV--1~TeV). No statistically significant emission is detected at the cluster position in any energy band. In the hard X-ray regime, we derive a $3\sigma$ upper limit of $F_{30-100\,\mathrm{keV}} \lesssim 8.5 \times 10^{-11}$~erg~cm$^{-2}$~s$^{-1}$ from ISGRI mosaic imaging. Reanalysis of archival COMPTEL data yields 95\% confidence-level upper limits ranging from $\sim 9 \times 10^{-11}$ to $\sim 8 \times 10^{-10}$~erg~cm$^{-2}$~s$^{-1}$ across 0.75--30~MeV. In the GeV--TeV range, DAMPE \rthis{places upper limits on the}   differential energy flux to $\sim 10^{-12}$--$10^{-10}$~erg~cm$^{-2}$~s$^{-1}$ (95\% confidence level). These results provide independent multi-band constraints on the reported GeV excess from recent Fermi-LAT studies. While our DAMPE limits do not exclude the flux levels claimed in those analyses, the absence of confirmation across keV--MeV--GeV bands indicates that any non-thermal emission from Abell~119 remains tentative. 
\end{abstract}


\maketitle
\section{\label{sec:level1}Introduction\protect}
Galaxy clusters are the most massive gravitationally bound structures in the Universe and can be used to probe cosmology~\citep{White1978,Kravtsov2012,Allen2011,Vikhlinin2014} as well as fundamental physics~\citep{Desai2018,Bora2021,Bora2021b,Bora2022}. Approximately 80\% of the mass in these clusters is attributed to dark matter, while the intracluster medium (ICM) consisting of hot diffuse gas accounts for about 10-15\% of their total mass~\citep{Murase2013,Condorelli2023}.  Although galaxy clusters have been been detected across the electromagnetic spectrum  from radio  to X-rays, an unequivocal detection of gamma rays remains an open question. A summary of recent results  in the searches for gamma-ray emission from individual galaxy clusters as well as stacked emission  can  be found in ~\citet{Baghmanyan2022,Keshet,Mannastacked,Judit,Li2026} and (references therein).

In one of our recent works~\citep{Manna2024}, we carried out a systematic search for gamma-rays from 300 galaxy clusters selected from the 2500 sq. degree SPT-SZ survey~\citep{Bleem15,Bocquet19} based on their values $M_{500}/z^2$ using 15 years of Fermi-LAT data.  From this search, 
we were able to detect a gamma-ray signature from one galaxy cluster (namely, SPT-CL J2012-5649) with a significance of $6.1\sigma$~\citep{Manna2024}. The signal was detected in the energy range between 1-10 GeV with a spectral index of approximately -3.6. However, given the  point-spread function of Fermi-LAT, we could not ascertain if this signal is due to the intracluster medium or from radio galaxies in the vicinity of the cluster. This signal has been subsequently confirmed by other works~\citep{Li2026}.  We then searched for this signal in DArk Matter Particle Explorer (DAMPE) gamma-ray detector (between 3 GeV-1 TeV)~\citep{Manna2024c}, COMPTEL (between 0.75-30 MeV)~\citep{Manna2024b} and INTEGRAL/ISGRI (between 30-300 keV)~\citep{Manna2025i}. However, null results were obtained for searches from all these detectors. 

Another cluster for which tentative evidence for  gamma-ray emission has recently been reported is Abell~119~\citep{Harale2025,Li2026}.  This cluster is a nearby massive galaxy cluster ($z = 0.044$)~\citep{Smith2004} with an estimated mass of $M \approx 3.05 \times 10^{14}\, M_\odot$~\citep{Way1997}. It is characterized by a high X-ray luminosity of approximately $3 \times 10^{44}$~erg~s$^{-1}$, indicative of a dense and thermally energetic intracluster medium~\citep{Lee2016,Markevitch1998}. Multiwavelength observations reveal that Abell~119 is dynamically disturbed, exhibiting evidence of past and possibly ongoing off-axis merger events~\citep{Watson2023}. 
In a dedicated study based on 14 years 
of observations, \citet{Harale2025} identified a $\sim 4\sigma$ excess of 
diffuse emission offset by $\sim 0.25^{\circ}$ from the cluster center, 
spatially overlapping with its virial region. Using spatially extended 
templates and various cosmic-ray proton (CRp) distribution models, they 
obtained Test Statistic (TS) values of $\sim 18$, and derived 
integrated $\gamma$-ray fluxes in the 100~MeV--1~TeV band of 
$(10$--$12) \times 10^{-10}$~ph~cm$^{-2}$~s$^{-1}$. The best-fit spectrum 
is characterized by a photon spectral index of $\Gamma \approx 2.2$, corresponding to an inferred $\gamma$-ray luminosity of $\sim 1.2 \times 10^{43}$~erg~s$^{-1}$. Interpreting the signal within a hadronic framework, they estimated a cosmic-ray proton-to-thermal energy fraction of $\sim 7$--$8\%$ and predicted an associated neutrino flux potentially detectable by next-generation observatories. However, the TS values remain below the canonical $5\sigma$ 
detection threshold (TS=25), and the authors cautioned that the signal does not yet constitute a firm detection.
Around the same time, \rthis{\citet{Li2026} conducted a systematic search for $\gamma$-ray emission from 65 HIFLUGCS galaxy clusters using 16 years of Fermi-LAT data. Among these clusters, Abell~119 exhibited a moderate excess with TS~$\approx 20.7$ after optimized source localization, 
corresponding to an integrated flux of $(2.53 \pm 0.49) \times 10^{-10}$~ph~cm$^{-2}$~s$^{-1}$  in the 0.5--500~GeV band. This significance remained below the formal detection threshold of 5$\sigma$. The authors identified five sources from 
the PMN catalog~\citep{Wright1994} within the 95\% confidence region of the $\gamma$-ray position, with two sources, PMN J0056-0116 and PMN J0056-0100, located within $0.2^\circ$ of the $\gamma$-ray position. Radio-loud active galactic nuclei (AGN), particularly those associated with relativistic jets, 
are well-established sources of GeV $\gamma$-ray emission 
\citep{Ghirlanda2011}. The presence of these nearby radio sources therefore provides a plausible alternative point-source contribution to the reported GeV excess, although their spatial coincidence alone does not establish that they are responsible for the observed emission. This 
possibility motivates independent observations with complementary $\gamma$-ray instruments such as DAMPE.} For the clusters with TS~$> 16$, 
dark matter annihilation scenarios were also found to be incompatible with 
constraints from dwarf spheroidal galaxies, favoring instead a hadronic 
cosmic-ray origin for the observed emission.

Prior searches for non-thermal emission from Abell~119 at hard X-ray energies 
have yielded only upper limits. A  systematic analysis 
of HIFLUGCS galaxy clusters  was conducted using joint data from the Swift/BAT all-sky survey data 
and XMM-Newton EPIC observations spanning 2--195~keV, searching for 
excess emission above the thermal intracluster medium component~\citep{Wik2012}. For 
Abell~119, their joint spectral fits produced a 90\% confidence level upper limit 
of $1.69 \times 10^{-12}$~erg~cm$^{-2}$~s$^{-1}$ on the 20--80~keV non-thermal 
flux (including systematic uncertainties). Assuming a fixed power-law index 
of $\Gamma = 2$, they derived a $3\sigma$ upper limit of 
$6.58 \times 10^{-12}$~erg~cm$^{-2}$~s$^{-1}$. 
Allowing the spectral index to 
vary yielded a best-fit value of $\Gamma \sim 2.12$ and a corresponding 
$3\sigma$ limit of $6.43 \times 10^{-12}$~erg~cm$^{-2}$~s$^{-1}$. These 
results indicate that there is no statistically significant excess at hard X-ray energies  above the  thermal emission from the cluster.

Despite the recent  reported GeV excess from Abell 119, the statistical significances remain below the formal $5\sigma$ detection threshold, and the origin of the emission remains uncertain. In particular, no confirmation has yet been obtained across independent instruments or complementary energy bands. While hard X-ray upper limits exist, no coordinated broadband analysis has tested the spectral consistency of the non-thermal interpretation from the keV to TeV regime for Abell~119. Such a multi-instrument study is required to assess the physical plausibility of the claimed excess and to place robust constraints on the cosmic-ray energy content of the cluster.

In this work, we search for high-energy emission from Abell~119 over a broad energy range spanning hard X-rays to  TeV energies using archival observations from INTEGRAL/ISGRI, COMPTEL, and DAMPE. The structure of this manuscript is as follows. The data analysis procedures for each instrument are described in Sect.~\ref{sec:data_analysis}. The results are presented in Sect.~\ref{sec:results}. Finally, we summarize our findings in conclusions in Sect.~\ref{sec:conclusions}.

\section{Data Analysis}
\label{sec:data_analysis}
We perform a search for non-thermal emission from Abell~119~\citep{Harale2025} using observations spanning keV to TeV energies. The analysis combines data from INTEGRAL/ISGRI~\citep{Ubertini2003,Winkler2003}, COMPTEL~\citep{Schoenfelder1993}, and DAMPE~\citep{Chang2014,Chang2017,Ambrosi2019}, providing energy coverage from $\sim$30~keV to 1~TeV. Each instrument was analyzed independently using its standard data reduction and statistical framework.
For instruments employing likelihood-based analyses, the detection significance was quantified using the Test Statistic (TS), defined as
\begin{equation}
\mathrm{TS} = -2 \ln \left( \frac{\mathcal{L}_0}{\mathcal{L}_1} \right),
\label{eq:TS}
\end{equation}
where $\mathcal{L}_0$ and $\mathcal{L}_1$ denote the maximum likelihood values under the null (background-only) and source-included hypotheses, respectively. Under Wilks' theorem~\citep{Wilks1938}, TS asymptotically follows a half-$\chi^2$ distribution with degrees of freedom equal to the number of additional free parameters in the source model~\citep{Mattox96}. One can then show  that $\sqrt{\mathrm{TS}}$ approximates the detection significance in Gaussian standard deviations for one degree of freedom. In the absence of significant excess emission, 95\% confidence-level flux upper limits were derived. The instrument-specific analysis procedures are described below.

\subsection{INTEGRAL/ISGRI Data Analysis}
\label{sec:isgri_analysis}
The search for hard X-ray and soft $\gamma$-ray emission from Abell~119 was conducted using the ISGRI detector of the IBIS instrument~\citep{Ubertini2003} onboard INTEGRAL~\citep{Winkler2003}. All publicly available Science Windows (ScWs) covering the position of Abell~119 (RA~=~$14.035^\circ$, Dec~=~$-1.200^\circ$, J2000) were retrieved from the INTEGRAL Science Data Archive~\footnote{\url{https://www.isdc.unige.ch/integral/archive}}. After applying standard quality filters to exclude time intervals affected by solar flares and instrumental anomalies, a total of 38 Science windows were identified within a $5^\circ$ radius of the cluster center (following the approach in~\citet{Manna2025i}). All data were processed using the INTEGRAL Off-line Scientific Analysis software package (OSA v11.2)~\citep{Curvoisier2003}, following standard procedures outlined in the OSA User Manual\footnote{\url{https://www.astro.unige.ch/integral/analysis}} and consistent with our previous INTEGRAL analysis~\citep{Manna2025i}.

The image reconstruction was performed from the energy-corrected level (\texttt{startLevel = COR}) through the image reconstruction level (\texttt{endLevel = IMA2}). The default ISGRI reference catalog (\texttt{CAT\_refCat = \$ISDC\_REF\_CAT[ISGRI\_FLAG>0]}) was adopted during processing. The imaging analysis was configured to produce sky mosaics in four logarithmically spaced energy bands: 30--41, 41--54.5, 54.5--73.5, and 73.5--99.5~keV, similar to our previous work~\citep{Manna2025i}. These bands were chosen to optimize sensitivity given the energy-dependent effective area of ISGRI while maintaining adequate photon statistics. The lower bound of 30~keV reflects the degraded low-energy response of ISGRI after Revolution~1600 (mid-2015), as documented in the OSA User Manual~\citep{Curvoisier2003}, while the upper bound of $\sim$100~keV avoids the rapidly declining effective area at higher energies.

Sky images were generated using the \texttt{ii\_skyimage} task. Source detection employed a combined search approach (\texttt{OBS1\_SearchMode = 3}) that includes catalog sources and up to 50 additional candidates, with a minimum detection significance threshold of $5\sigma$ for newly identified sources. The background subtraction was performed using the standard ISGRI background maps provided by the IBIS team. A mosaic image was constructed by combining all 38 ScWs with pixel spreading enabled (\texttt{OBS1\_PixSpread = 1}) to improve the source localization accuracy.

The statistical significance at the source position was obtained directly from the ISGRI mosaic significance maps produced by \texttt{ii\_skyimage}, which provide the significance in units of Gaussian standard deviations. We adopt a $5\sigma$ threshold (corresponding to a TS value of 25) as the criterion for a robust detection, consistent with standard practice in high-energy astrophysics~\citep{Lyons2013}. 

\subsection{COMPTEL Data Analysis}
\label{sec:comptel_analysis}

We now analyze archival COMPTEL observations to probe potential emission in the 0.75--30~MeV range, thereby bridging the gap between hard X-rays/soft gamma-rays and high-energy $\gamma$-rays.

The search for MeV $\gamma$-ray emission from Abell~119 was conducted using archival data from the Imaging Compton Telescope (COMPTEL) aboard the Compton Gamma Ray Observatory (CGRO), which operated from April 1991 through June 2000~\citep{Schoenfelder1993}. COMPTEL was sensitive in the 0.75--30~MeV energy range with a field of view of approximately one steradian, achieving an energy resolution of 5--8\% and an angular resolution of $(1.7$--$4.4)^{\circ}$ (FWHM), depending on energy~\citep{Schoenfelder1993}. To date, no subsequent mission has provided comparable all-sky imaging sensitivity in the MeV regime, making the COMPTEL archive uniquely valuable for studies of soft $\gamma$-ray emission from galaxy clusters. Recent developments in the GammaLib and ctools framework~\citep{Knodlseder2016,Knodlseder2022} have facilitated the reanalysis of COMPTEL data. The data reduction and analysis procedures follow the methodologies detailed in our previous COMPTEL studies of galaxy clusters~\citep{Manna2024b} and red dwarfs~\citep{Shrivastava2024} which made use of these recent advancements, with adaptations specific to Abell~119. We summarize the key steps below.

We queried the COMPTEL archive to identify all viewing periods (VPs) that included Abell~119 within a $30^{\circ}$ radius of the pointing direction, similar to~\citet{Shrivastava2024}, resulting in four suitable VPs. The \texttt{comobsselect} utility was used to extract the corresponding observations. To enable energy-resolved analysis, the \texttt{comobsbin} tool organized the data into logarithmically spaced energy bins spanning 0.75--30~MeV. Two binning configurations were employed: a fine scheme with 16 energy bins optimized for spectral characterization, following~\citet{Manna2024b,Shrivastava2024}, and a coarser 4-bin scheme designed to maximize photon statistics per bin, consistent with standard COMPTEL analysis practice\footnote{\url{http://cta.irap.omp.eu/ctools/users/tutorials/comptel/index.html}}. 

COMPTEL data are analyzed in a three-dimensional Compton data space defined by the reconstructed scatter angle ($\bar{\varphi}$) and two scatter direction coordinates ($\chi$, $\psi$)~\citep{Knodlseder2022}. Multiple viewing periods were merged using the \texttt{comobsadd} tool with the standard COMPTEL binning configuration: 80 bins with $1^{\circ}$ spacing in each scatter direction coordinate and 25 bins with $2^{\circ}$ spacing in scatter angle. The combined effective exposure from all four viewing periods totals $7.23 \times 10^{9}$~cm$^{2}$~s at the position of Abell~119. Model definition files were generated using \texttt{comobsmodel} for maximum-likelihood fitting. Abell~119 was modeled as a point source with a power-law spectrum characterized by normalization $N_0$ and photon spectral index $\Gamma$. Standard COMPTEL instrumental background components were included in the likelihood model as implemented within GammaLib.

Spectral analysis was performed using the \texttt{csspec} tool. We adopted the \texttt{BINS} method, which replaces the assumed spectral model with independent normalization parameters in each energy bin, providing a model-independent approach particularly suitable for weak sources or non-detections. For each energy bin, the likelihood fit yields the flux estimate, statistical uncertainty, TS values, and 95\% confidence level upper limits. The detection significance was evaluated using TS defined in Eq.~\ref{eq:TS}. 

\subsection{DAMPE Data Analysis}
\label{sec:dampe_analysis}
Next, we analyzed observations from the DAMPE, which provides sensitivity to $\gamma$ rays between 3~GeV and 10~TeV~\citep{Chang2017,Ambrosi2019,Chang2014}. The data reduction and analysis procedures closely follow the methodology detailed in our previous DAMPE study of the SPT-CL J2012-5649 cluster~\citep{Manna2024c}, with adaptations specific to Abell~119. We summarize the key steps below.

DAMPE achieves an angular resolution of $\leq 0.2^{\circ}$ at 100~GeV with a field of view of approximately 1.0~sr and provides an excellent energy resolution of $\sim$1\% at 100~GeV~\citep{Chang2017}. This advantage comes at the cost of a smaller acceptance, roughly seven times lower than that of Fermi-LAT. Detailed descriptions of the DAMPE detector subsystems and in-orbit performance calibrations can be found in ~\citet{Chang2017,Chang2014,Ambrosi2019}.

We analyzed DAMPE photon data collected between January~1,~2016 and December~29,~2023, corresponding to the Mission Elapsed Time \texttt{MET 94608000--346809600}. The analysis was carried out using the \texttt{DAMPE Science Tools (DmpST)} software package~\citep{Duan}, which combines photon event data, spacecraft pointing information, Monte Carlo instrument response functions (IRFs), and likelihood-based analysis tools. Photon events were selected within a circular region of interest (ROI) of radius $10^{\circ}$ centered on the cluster position. The 10$^\circ$ ROI ensures adequate modeling of diffuse backgrounds while encompassing the instrument point-spread function at low energies, and is similar to that adopted in~\citet{Manna2024c}. \rthis{The energy range of the analysis was restricted to 3~GeV--1~TeV, following our previous DAMPE analysis of galaxy clusters~\citep{Manna2024}. This range provides the relevant overlap with the GeV emission reported for Abell~119 while retaining adequate photon statistics for a likelihood analysis of this faint source. Although DAMPE is sensitive to $\gamma$ rays up to energies of order 10~TeV~\citep{Chang2017,Ambrosi2019}, the number of detected photons decreases with increasing energy, leading to progressively poorer photon statistics and reduced sensitivity at the highest energies, as also illustrated by the DAMPE sensitivity curve. We therefore restrict the present analysis to 1~TeV, both to maintain consistency with our previous cluster analysis and to focus on the energy range most relevant for testing the reported GeV excess.} We retained only events satisfying the High-Energy Trigger (HET) selection criteria. Data recorded during passages through the South Atlantic Anomaly, periods affected by intense solar activity, and times when the target was more than $60^{\circ}$ from the spacecraft $Z$-axis were excluded. The total livetime for the Abell~119 ROI is approximately $1.92 \times 10^{8}$~s.

Spatial binning was performed with a pixel size of $0.1^{\circ}$, while the energy range was divided into ten logarithmically spaced bins between 3~GeV and 1~TeV~\citep{Manna2024c}. The source model consisted of a point-source component located at the cluster center, described by a power-law spectral model with spectral index initially fixed to $\Gamma = 2$, consistent with previous studies of galaxy clusters~\citep{Baghmanyan2022,Manna2024}. Diffuse $\gamma$-ray backgrounds were modeled using the standard Galactic diffuse emission template \texttt{gll\_iem\_v07.fits} together with an isotropic component represented by a power-law spectral model. The normalizations of both diffuse components were treated as free parameters during the likelihood fitting procedure. A standard binned maximum-likelihood analysis was performed using the \texttt{DmpST} likelihood framework to determine the best-fit spectral parameters. The significance of the emission was also quantified using TS (cf. Eq.~\ref{eq:TS}), similar to the other detectors.

\section{Results}
\label{sec:results}
In the following subsections, we present the results of our multi-instrument search for non-thermal emission from Abell~119. No statistically significant excess($>5\sigma$) was found in any of the analyzed energy bands, and we therefore obtain upper limits as presented in subsequent subsections.  

\subsection{INTEGRAL/ISGRI Hard X-ray Analysis}
\label{sec:isgri_results}

\subsubsection{Source Detection and Catalog Cross-matching}
\begin{figure*}
\centering
\includegraphics[width=0.48\textwidth]{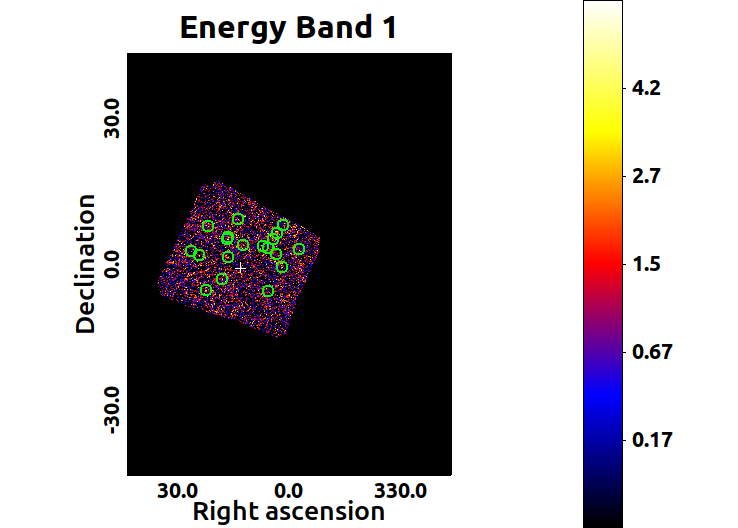}
\includegraphics[width=0.48\textwidth]{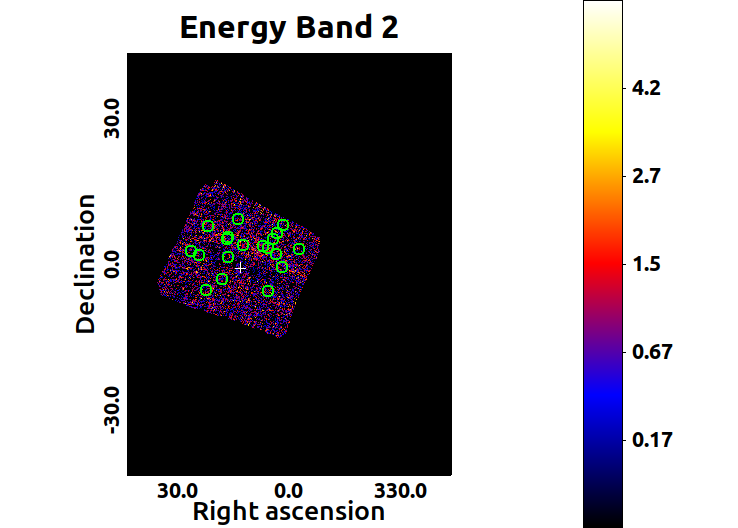}
\includegraphics[width=0.48\textwidth]{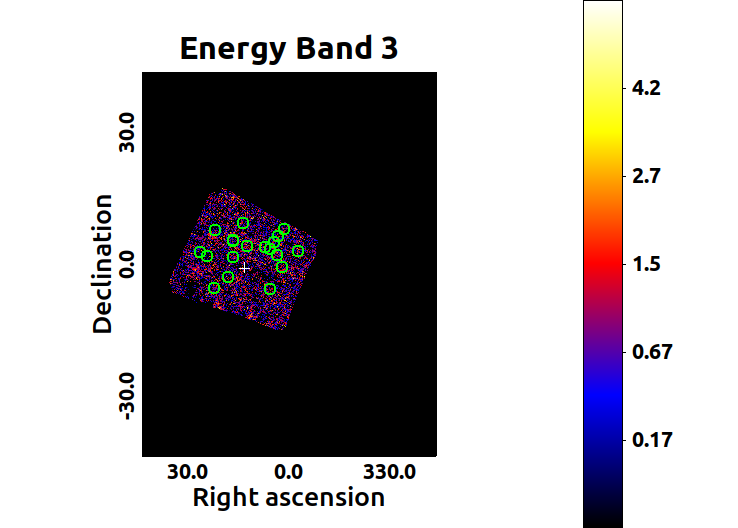}
\includegraphics[width=0.48\textwidth]{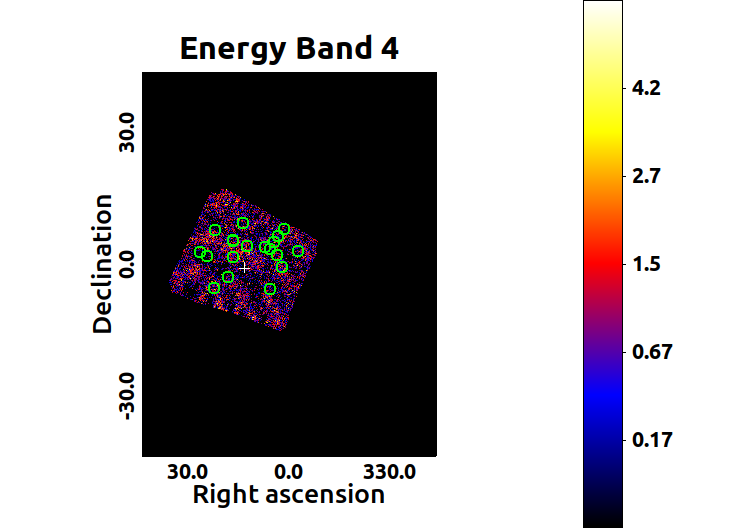}
\caption{INTEGRAL/ISGRI mosaic significance maps of the Abell~119 field in four logarithmically spaced energy bands: Band~1 (30--41~keV; \emph{top left}), Band~2 (41--54.5~keV; \emph{top right}), Band~3 (54.5--73.5~keV; \emph{bottom left}), and Band~4 (73.5--99.5~keV; \emph{bottom right}). The white cross marks the cluster center at RA~=~$14.035^{\circ}$, Dec~=~$-1.200^{\circ}$ (J2000). Green circles indicate sources detected at $>5\sigma$ significance in the ISGRI mosaic images. The color scale ranges from $0\sigma$ (blue) to $+6\sigma$ (red). No significant emission is detected at the position of Abell~119 in any energy band.}
\label{fig:significance_maps}
\end{figure*}

The ISGRI mosaic images reveal multiple statistically significant point sources across the large field of view ($\sim$29$^{\circ}$ × 29$^{\circ}$)~\citep{Lebrun2003}. Figure~\ref{fig:significance_maps} presents the significance maps in all four energy bands. We identified a total of 28 sources within the analyzed region above the significance threshold of $5\sigma$ with 9 of those sources matching entries in the IBIS/ISGRI 1000-orbit source catalog~\citep{Bird2016} and 19 previously uncatalogued sources (marked with green circles in Figure~\ref{fig:significance_maps}). The newly detected sources likely correspond to previously uncatalogued hard X-ray emitters; detailed identification is beyond the scope of this work.

\subsubsection{Search for Emission from Abell~119}

To assess the potential association between the detected source candidates and
Abell~119, we computed their angular separations from the cluster center
(RA~$=14.035^{\circ}$, Dec~$=-1.200^{\circ}$, J2000). Table~\ref{tab:a119_isgri_all_sources}
lists the 28 source candidates identified in the ISGRI field, comprising
9 sources matching entries in the IBIS/ISGRI 1000-orbit source
catalog~\citep{Bird2016} and 19 previously uncatalogued candidates.
The detection significance of each source candidate is reported in units
of Gaussian standard deviations ($\sigma$), as given by the
\texttt{DETSIG} parameter in the OSA ISGRI source-results product.
The candidates are sorted by angular separation from the cluster center.
The nearest source candidate lies at an angular separation of
$\sim$290$'$ ($\sim$4.8$^{\circ}$), far beyond the ISGRI point spread
function (PSF; FWHM $\sim$12$'$)~\citep{Ubertini2003,Winkler2003}.
Thus, none of the identified source candidates is spatially coincident
with the position of Abell~119.

\begin{table*}
\centering
\caption{\rthis{ISGRI source candidates in the Abell~119 field, sorted by angular separation from the cluster center (RA = $14.035^{\circ}$, Dec = $-1.200^{\circ}$, J2000). The table lists both cataloged sources~\citep{Bird2016} and previously uncatalogued source candidates identified in this work. The detection significance is given in units of Gaussian standard deviations ($\sigma$), as reported by the OSA ISGRI source-results product. All source candidates lie well outside the ISGRI PSF (FWHM $\sim$12$'$)~\citep{Ubertini2003}}}
\label{tab:a119_isgri_all_sources}
\begin{ruledtabular}
\begin{tabular}{lccccr}
Source Name & Type & RA (deg) & Dec (deg) & Separation (arcmin) & Significance ($\sigma$) \\
\hline
NEW\_11            & New      & 17.5343 & 2.1175      & 289.3  & 6.40 \\
NEW\_13            & New      & 19.1219 & $-3.9904$   & 347.8  & 5.30 \\
NEW\_7             & New      & 13.2397 & 5.6958      & 416.5  & 5.85 \\
NEW\_12            & New      & 17.7180 & 7.1789      & 549.0  & 4.76 \\
NEW\_10            & New      & 7.6001  & 5.4461      & 554.7  & 5.09 \\
NEW\_17            & New      & 17.5749 & 7.5801      & 567.8  & 5.86 \\
NEW\_14            & New      & 5.9893  & 4.8192      & 602.5  & 4.70 \\
NEW\_8             & New      & 6.0631  & $-7.9744$   & 626.3  & 5.84 \\
NEW\_15            & New      & 23.1003 & $-7.0785$   & 646.9  & 6.27 \\
NEW\_3             & New      & 3.9155  & 2.9055      & 655.1  & 6.38 \\
NEW\_19            & New      & 24.8955 & 2.4303      & 686.9  & 4.93 \\
NEW\_1             & New      & 2.1170  & $-0.8284$   & 715.3  & 6.15 \\
NEW\_6             & New      & 4.7785  & 7.3055      & 753.3  & 4.01 \\
NEW\_4             & New      & 26.7931 & 3.6057      & 817.6  & 5.24 \\
NEW\_9             & New      & 14.6750 & 12.7249     & 836.4  & 5.22 \\
NEW\_16            & New      & 22.7021 & 10.2784     & 861.5  & 6.07 \\
NEW\_2             & New      & 3.3959  & 9.0909      & 886.4  & 5.03 \\
NEW\_18            & New      & 1.6561  & 11.4312     & 1058.0 & 5.25 \\
NEW\_5             & New      & 356.8664 & 4.3916     & 1082.6 & 5.64 \\
\hline
Mrk~1152           & Catalog  & 18.4587  & $-14.8456$ & 859.7  & 0.00 \\
IGR~J01528$-$0326  & Catalog  & 28.2042  & $-3.4468$  & 860.0  & 8.48 \\
IGR~J01528$-$0845  & Catalog  & 28.2340  & $-8.7690$  & 962.0  & 3.71 \\
Mrk~584            & Catalog  & 30.0990  & 2.7110     & 991.7  & 3.55 \\
NGC~788            & Catalog  & 30.2769  & $-6.8155$  & 1028.5 & 5.32 \\
Mrk~1018           & Catalog  & 31.5667  & $-0.2914$  & 1053.2 & 2.70 \\
IGR~J23558$-$1047  & Catalog  & 358.9951 & $-10.7791$ & 1064.8 & 3.25 \\
NGC~7603           & Catalog  & 349.7359 & 0.2435     & 1460.4 & 0.00 \\
PKS~2325$+$093     & Catalog  & 351.9370 & 9.6660     & 1472.4 & 0.00 \\
\end{tabular}
\end{ruledtabular}
\end{table*}

We separately quantified the detection significance at the exact position
of Abell~119 by extracting the corresponding values from the ISGRI mosaic
significance maps in each energy band. The maximum significance at the
cluster position is $1.21\sigma$, occurring in the 30--41~keV band.
Table~\ref{tab:a119_significance} presents the significance values in all
four energy bands. In every band, the significance at the cluster center
remains well below the canonical $5\sigma$ detection threshold, with
values ranging from 0 to $1.21\sigma$. These values are consistent with
background fluctuations and provide no evidence for hard X-ray or soft
$\gamma$-ray emission associated with Abell~119. Based on this
non-detection, we derive a $3\sigma$ upper limit on the hard X-ray flux
from Abell~119, as detailed below.

\begin{table}[h!]
\centering
\caption{Detection significance at the Abell~119 cluster center extracted from ISGRI mosaic significance maps. All values are consistent with background fluctuations.}
\label{tab:a119_significance}
\begin{ruledtabular}
\begin{tabular}{cc}
Energy Range & Significance \\
(keV) & ($\sigma$) \\
\hline
30--41       & 1.21 \\
41--54.5     & 0.00 \\
54.5--73.5   & 0.40 \\
73.5--99.5   & 0.00 \\
\end{tabular}
\end{ruledtabular}
\end{table}

\subsubsection{Flux Upper Limits}
\label{subsec:isgri_upperlimits}

Given the non-detection of Abell~119 in all the  ISGRI energy bands, we derive conservative flux upper limits using a purely image-based approach that avoids the systematic uncertainties associated with spectral fitting of coded-mask telescope data, as done in~\citet{Krivonos2010,Manna2025i}. We focus on the combined 30--100~keV band to maximize photon statistics while remaining within the optimal energy range of ISGRI, before the rapid decline of effective area above $\sim$100~keV.

The statistical uncertainty at the cluster position was extracted directly from the mosaic variance map, which encodes the photon counting statistics and systematic uncertainties from the imaging reconstruction. The $1\sigma$ count-rate uncertainty is given by
\begin{equation}
\sigma_{\rm cnt} = \sqrt{V},
\end{equation}
where $V$ is the variance value at the pixel corresponding to the cluster coordinates. For Abell~119, we obtain
\begin{equation}
\sigma_{\rm cnt} = 0.193~\mathrm{counts~s^{-1}}.
\end{equation}
Adopting a conservative $3\sigma$ confidence level appropriate for upper limit determinations in non-detections, the count rate is
\begin{equation}
C_{3\sigma} = 3\,\sigma_{\rm cnt} = 0.579~\mathrm{counts~s^{-1}}.
\end{equation}

To convert this count-rate upper limit into a physical energy flux, we employ an empirical calibration based on contemporaneous ISGRI observations of the Crab Nebula, processed through an identical analysis pipeline with matching energy bands and mosaic parameters~\citep{Krivonos2010,Manna2025i}. This approach accounts for the energy-dependent instrumental response, effective area, and imaging systematics. \rthis{We adopt the standard Crab spectrum commonly used for IBIS/ISGRI
calibration in the hard X-ray band~\citep{Krivonos2010,Krivonos2022}:
\begin{equation}
N(E) = 10\left(\frac{E}{1~\mathrm{keV}}\right)^{-2.1}
\;\mathrm{ph~cm^{-2}~s^{-1}~keV^{-1}},
\end{equation}
corresponding to  $\Gamma=2.1$. We use this spectral
prescription over the 30--100~keV band to calculate the integrated
Crab energy flux and the corresponding ISGRI count rate. The resulting
count-rate-to-flux calibration factor is then applied to the measured
upper limit for Abell~119. We note that
\citet{Kirsch2005} highlighted systematic differences between
instruments and the increasing complexity of the Crab spectrum above
$\sim30$~keV, making the Crab an imperfect absolute calibration
standard in this regime. Our analysis, however, uses the Crab only as
an empirical count-rate-to-flux calibration for the 30--100~keV
ISGRI mosaic, rather than relying on a detailed spectral fit to the
Crab or to Abell~119. The adopted Crab prescription is also
consistent with that used in the 17-year INTEGRAL/IBIS survey of
\citet{Krivonos2022}. We therefore retain this calibration
prescription while restricting its application to the 30--100~keV
band. The resulting calibration factor is then applied to the measured
upper limit for Abell~119.}

This procedure yields a $3\sigma$ upper limit on the energy flux from Abell~119 of
\begin{equation}
F_{3\sigma}(30\text{--}100~\mathrm{keV})
= 8.5 \times 10^{-11}\;
\mathrm{erg\,cm^{-2}\,s^{-1}}.
\end{equation}

For comparison with previous INTEGRAL studies of galaxy clusters and other faint hard X-ray sources, this limit corresponds to approximately 8--10~mCrab, where 1~mCrab $\sim$ (0.8--1.0)$\times10^{-11}$~erg~cm$^{-2}$~s$^{-1}$ in the 20--100~keV range~\citep{Bird2016}. This provides a stringent constraint on soft $\gamma$-ray emission from Abell~119 with INTEGRAL/ISGRI. The model-independent, image-based methodology employed here avoids potential systematic biases from spectral fitting procedures and source confusion that can affect coded-mask telescope analyses in crowded fields. 

\subsection{COMPTEL MeV Gamma-Ray Analysis}
\label{sec:comptel_results}

\subsubsection{Maximum-Likelihood Spectral Analysis}

Two independent likelihood analyses were performed with complementary energy binning strategies to test for potential emission while maximizing statistical robustness. An initial fit over the full 0.75--30~MeV range using 16 logarithmically spaced energy bins treated both $N_0$ and $\Gamma$ as free parameters, with the source position and $E_0$ held fixed. This configuration yielded $\mathrm{TS} = 0.61$, far below the detection threshold ($\mathrm{TS} > 25$ for $5\sigma$ significance). The best-fit normalization was consistent with zero within uncertainties, and the background model alone adequately reproduced the observed photon distribution across all energy bins.

To improve statistical stability and reduce sensitivity to energy-dependent systematic effects, we repeated the analysis using four broader energy bins corresponding to COMPTEL's native energy bands (see Table~\ref{tab:comptel_csspec}). An unconstrained fit with both spectral parameters free again produced $\mathrm{TS} \approx 0$. We then performed a constrained fit fixing the photon spectral index at $\Gamma = 2.0$, a value characteristic of hadronic $\gamma$-ray emission from cosmic-ray interactions \citep{Ackermann2014}. This analysis converged with $\mathrm{TS} \approx 0$ and showed excellent agreement between the model-predicted event count and the observed count, confirming that the background-only hypothesis provides a statistically adequate description of the data. Both analyses consistently indicate no statistically significant MeV $\gamma$-ray emission from Abell~119.

\subsubsection{Energy-Resolved Upper Limits}

\begin{figure}
    \centering
    \includegraphics[width=0.75\linewidth]{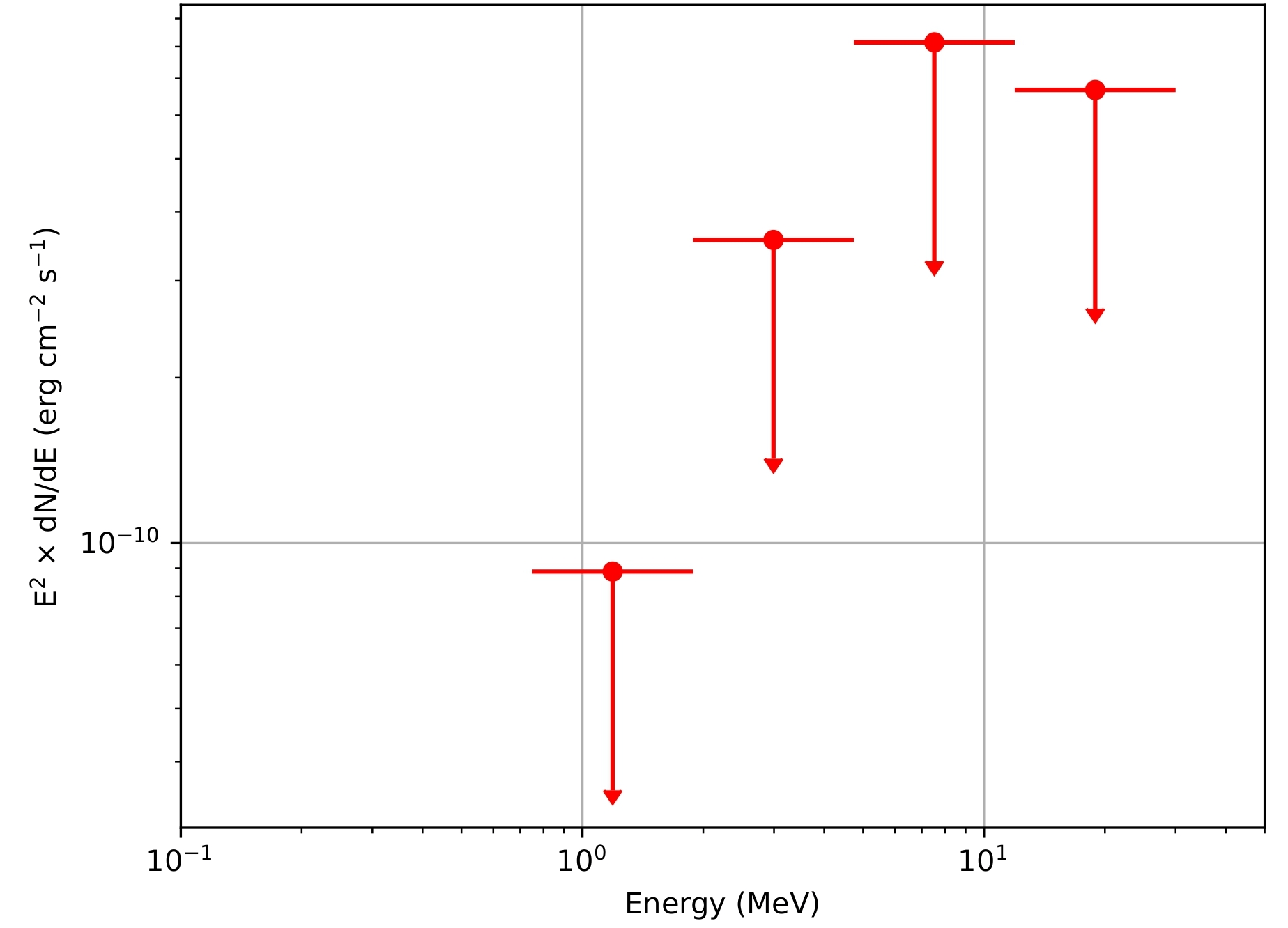}
    \caption{SED plot of Abell~119 in the MeV band from COMPTEL observations. The data points represent 95\% confidence-level upper limits on $E^{2}dN/dE$ derived from bin-by-bin maximum-likelihood analysis in four logarithmically spaced energy bins covering 0.75--30~MeV. No statistically significant $\gamma$-ray emission is detected.}
    \label{fig:sed2}
\end{figure}

Following the null detection in the likelihood analysis, we derived 95\% confidence-level upper limits in each of four logarithmically spaced energy bins spanning 0.75--30~MeV. Table~\ref{tab:comptel_csspec} summarizes the results. All energy bins yield TS~$\lesssim 2$, confirming the absence of detectable emission. The 95\% upper limits on the integrated energy flux range from $\sim 9 \times 10^{-11}$ to $\sim 8 \times 10^{-10}$~erg~cm$^{-2}$~s$^{-1}$, with the most stringent constraints obtained in the 0.75--1.89~MeV band. Figure~\ref{fig:sed2} presents the Spectral Energy Distribution (SED) plot showing upper limits across the full MeV range.

\begin{table}[htbp]
\centering
\caption{Energy-resolved 95\% confidence level upper limits on MeV $\gamma$-ray flux from Abell~119 derived from COMPTEL observations. The TS values in all bins are consistent with background fluctuations ($\mathrm{TS} \lesssim 2$), indicating no significant detection.}
\label{tab:comptel_csspec}
\begin{ruledtabular}
\begin{tabular}{ccc}
Energy Range & Flux Upper Limit & TS \\
(MeV) & (erg cm$^{-2}$ s$^{-1}$) & \\
\hline
0.75 -- 1.89  & $< 8.87 \times 10^{-11}$ & 0.01  \\
1.89 -- 4.74  & $< 3.56 \times 10^{-10}$ & 0.02 \\
4.74 -- 11.93 & $< 8.14 \times 10^{-10}$  & 1.97 \\
11.93 -- 30.00 & $< 6.67 \times 10^{-10}$ & 0.01 \\
\end{tabular}
\end{ruledtabular}
\end{table}

\subsection{DAMPE GeV--TeV Gamma-Ray Analysis}
\label{sec:dampe_results}

\subsubsection{Broadband Likelihood Analysis}
The broadband fit models potential emission from Abell~119 as a point source centered at the cluster coordinates (RA~=~$14.035^{\circ}$, Dec~=~$-1.200^{\circ}$, J2000) with a power-law spectrum. No statistically significant emission is detected from the cluster position. The best-fit Test Statistic is $\mathrm{TS} \simeq 0$, indicating that the inclusion of a source at the Abell~119 location does not improve the likelihood relative to the background-only hypothesis. This result is  consistent with statistical fluctuations. As expected for a non-detection, the fitted photon spectral index is poorly constrained. The diffuse background components dominate the photon counts within the ROI.

\subsubsection{Energy-Resolved Analysis and Upper Limits}

To investigate possible energy-dependent features, we performed a bin-by-bin likelihood analysis using ten logarithmically spaced energy bins spanning 3~GeV to 1~TeV, as also done in~\citet{Manna2024c}. In each bin, the source normalization was treated as a free parameter while fixing the photon spectral index to $\Gamma = 2.0$. 

\begin{figure}
    \centering
    \includegraphics[width=0.75\linewidth]{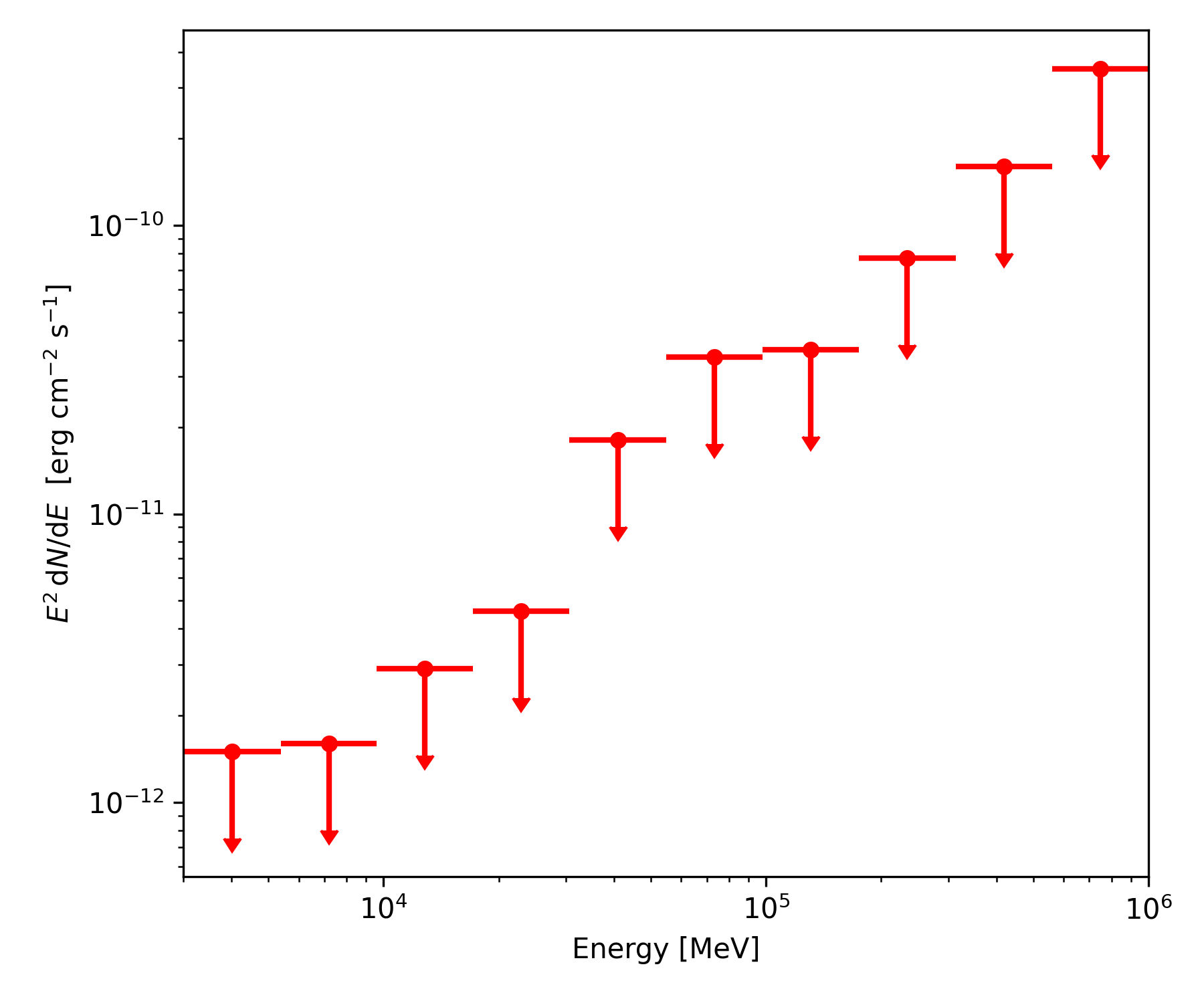}
    \caption{SED plot of Abell~119 from DAMPE observations. The data points represent 95\% confidence-level upper limits on $E^{2}dN/dE$ derived from bin-by-bin likelihood analysis in ten logarithmically spaced energy bins between 3~GeV and 1~TeV. No significant $\gamma$-ray excess is detected in any energy bin.}
    \label{fig:sed}
\end{figure}

In all DAMPE energy bins, the TS values remain far below the canonical detection threshold of TS~=~25, indicating the absence of any statistically significant $\gamma$-ray emission from Abell~119. We therefore derive bin-by-bin 95\% confidence-level upper limits on the energy flux across the full DAMPE energy range. The resulting SED plot is shown in Figure~\ref{fig:sed}, where the upper limits span nearly two orders of magnitude, from $\sim 10^{-12}$ to $\sim 10^{-10}$~erg~cm$^{-2}$~s$^{-1}$ between 3~GeV and 1~TeV. The numerical values of the upper limits, together with the corresponding TS values, are reported in Table~\ref{tab:dampe_sed}. \rthis{The absence of statistically significant emission across the entire
3~GeV--1~TeV band indicates that no GeV--TeV $\gamma$-ray excess from Abell~119 is detected above the sensitivity of the present DAMPE analysis.}

\begin{table}[htbp]
\centering
\caption{Energy-resolved 95\% confidence level upper limits on GeV--TeV $\gamma$-ray flux from Abell~119 derived from DAMPE observations. The TS values in all bins remain well below the detection threshold ($\mathrm{TS} < 25$), indicating no significant emission.}
\label{tab:dampe_sed}
\begin{ruledtabular}
\begin{tabular}{cccc}
Energy Range & Flux Upper limits & TS \\
(GeV) &  (erg cm$^{-2}$ s$^{-1}$) & \\
\hline
3.0 -- 5.4     & $< 1.5 \times 10^{-12}$ & 0.15 \\
5.4 -- 9.6     & $< 1.6 \times 10^{-12}$ & 0.06 \\
9.6 -- 17.1    & $< 2.9 \times 10^{-12}$ & 0.02 \\
17.1 -- 30.6   & $< 4.6 \times 10^{-12}$ & 0.01 \\
30.6 -- 54.8   & $< 1.8 \times 10^{-11}$ & 0.30 \\
54.8 -- 97.9   & $< 3.5 \times 10^{-11}$ & 0.07 \\
97.9 -- 175.0  & $< 3.7 \times 10^{-11}$ & 0.11 \\
175.0 -- 312.9 & $< 7.7 \times 10^{-11}$ & 0.03 \\
312.9 -- 559.4 & $< 1.6 \times 10^{-10}$ & 0.01 \\
559.4 -- 1000  & $< 3.5 \times 10^{-10}$ & 0.02 \\
\end{tabular}
\end{ruledtabular}
\end{table}

\subsection{Combined Analysis}
\label{subsec:combined_analysis}
To enable a uniform comparison of non-thermal constraints across multiple instruments and energy ranges, we performed a combined analysis of Abell~119 spanning hard X-rays to TeV energies, incorporating archival observations from Swift/BAT, INTEGRAL/ISGRI, COMPTEL, DAMPE, and Fermi-LAT. For energy-binned instruments (Swift/BAT, INTEGRAL/ISGRI, COMPTEL, and DAMPE), fluxes are plotted at the geometric mean energy of each bin, with downward arrows denoting upper limits. 

\begin{figure*}[t]
    \centering
    \includegraphics[width=0.9\textwidth]{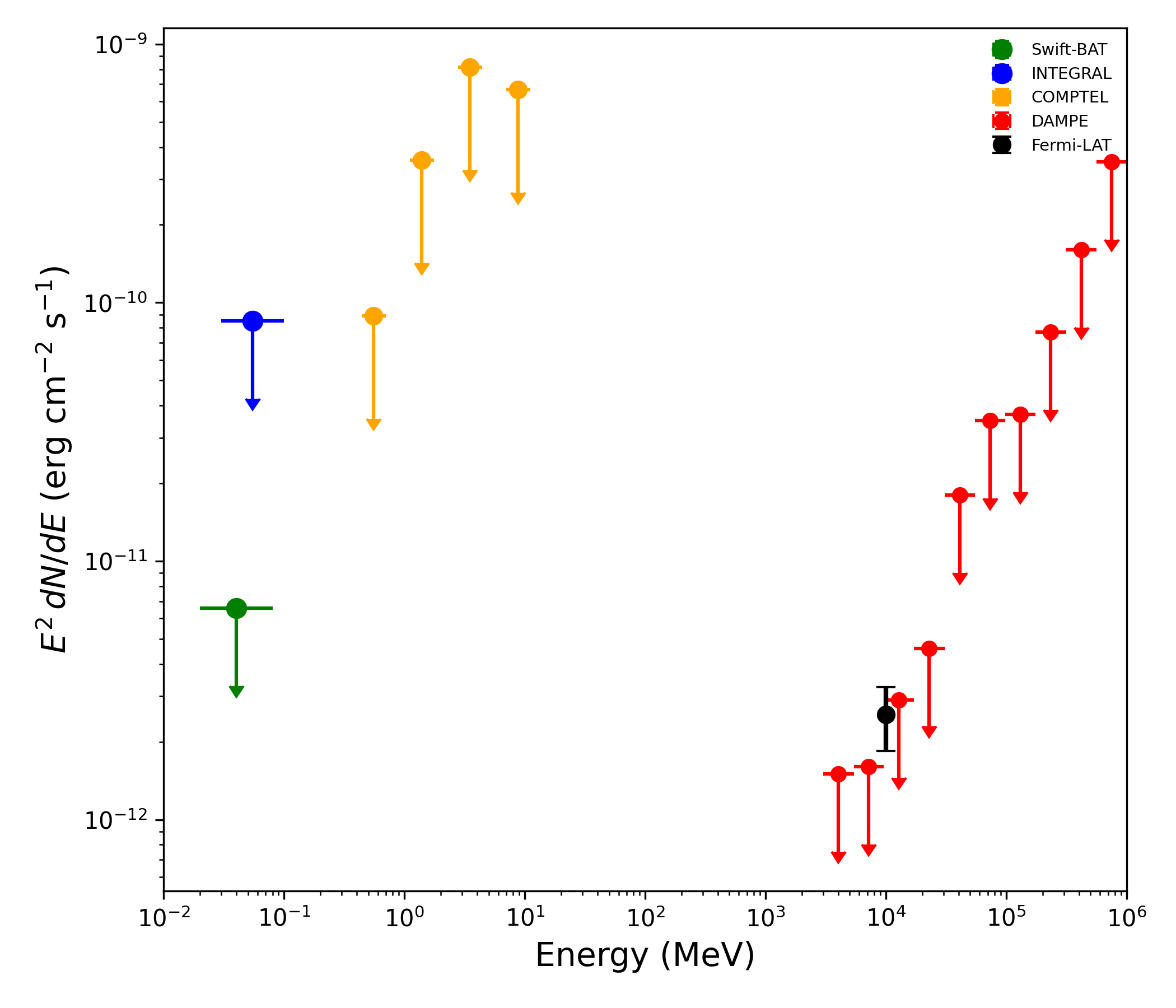}
    \caption{SED plot of Abell~119 from hard X-rays to GeV energies. Downward arrows with horizontal energy bars denote upper limits from Swift/BAT, INTEGRAL/ISGRI, COMPTEL, and DAMPE. The Fermi-LAT point shows the band-integrated energy flux for the maximum-TS spatial model of \citet{Harale2025}. }
    \label{fig:combined_sed}
\end{figure*}

For Swift/BAT and INTEGRAL/ISGRI, we adopted published $3\sigma$ upper limits on non-thermal emission in the 20--80~keV and 30--100~keV bands, respectively. COMPTEL upper limits were derived using the \texttt{csspec} framework within \texttt{ctools}, as shown in Table~\ref{tab:comptel_csspec}, and DAMPE upper limits were obtained from energy-resolved likelihood analyses as depicted in Table~\ref{tab:dampe_sed}. For the Fermi-LAT measurement, we adopted the spatial model yielding the maximum TS (TS~$= 18.47$; galaxy-density template)~\citep{Harale2025}. The associated band-integrated photon flux in the 100~MeV--1~TeV range was converted to an energy flux assuming a power-law spectrum with photon spectral index $\Gamma \approx 2.2$, and plotted at the geometric mean energy of the band. Figure~\ref{fig:combined_sed} presents the resulting SED of Abell~119, spanning over seven decades in energy. This combined representation enables a direct visual and quantitative assessment of the compatibility between the tentative GeV excess reported by \citet{Harale2025} and the non-thermal upper limits derived from hard X-ray and MeV--TeV observations.

\rthis{We also note that the different levels of the upper limits across the SED should not be interpreted as indicating a preference for emission in the MeV or TeV regimes. Rather, they primarily reflect the energy-dependent sensitivities of the individual instruments, which depend on factors such as effective area, exposure, photon statistics, instrumental and diffuse backgrounds, and angular and energy resolution. In particular, the comparatively weaker COMPTEL constraints reflect the challenging sensitivity of observations in the MeV regime, whereas DAMPE provides substantially tighter constraints over much of the GeV range. Thus, the upper limits constrain the allowed flux in each energy interval but do not by themselves identify an energy range in which emission is expected. Determining whether a physical non-thermal component is preferentially produced in the MeV, GeV, or TeV regime would require an assumed emission model and a corresponding broadband spectral analysis. This combined representation therefore enables a direct visual and quantitative assessment of the compatibility between the tentative GeV excess reported by \citet{Harale2025} and the non-thermal upper limits derived from hard X-ray and MeV--TeV observations.}

\section{Conclusions}
\label{sec:conclusions}
We have carried out a comprehensive broadband search for non-thermal high-energy 
emission from Abell~119 using archival observations from INTEGRAL/ISGRI (30--100~keV), COMPTEL (0.75--30~MeV), and DAMPE (3~GeV--1~TeV), providing coverage from hard X-rays to TeV $\gamma$-rays. 

In the hard X-ray band, INTEGRAL/ISGRI reveals no statistically significant emission in any  of the  energy bands (30--41, 41--54.5, 54.5--73.5, and 
73.5--99.5~keV), with all significances below $1.5\sigma$. We derive a $3\sigma$ 
upper limit on the integrated 30--100~keV flux of
\[
F_{30\text{--}100\,\mathrm{keV}} \lesssim 8.5 \times 10^{-11}~\mathrm{erg~cm^{-2}~s^{-1}},
\]
consistent with, though less stringent than, the 90\% confidence limit of $1.69 \times 10^{-12}$~erg~cm$^{-2}$~s$^{-1}$ (20--80~keV) from the joint Swift/BAT and XMM-Newton analysis~\citep{Wik2012}. Both studies independently exclude any significant hard X-ray excess above thermal intracluster emission, placing stringent constraints on inverse Compton or synchrotron emission from relativistic electrons in the cluster.

COMPTEL reanalysis yields no significant excess in broadband or energy-resolved fits (TS~$< 25$ throughout), with 95\% confidence level upper limits ranging from $\sim 9 \times 10^{-11}$~erg~cm$^{-2}$~s$^{-1}$ at 0.75--1.89~MeV to $\sim 8 \times 10^{-10}$~erg~cm$^{-2}$~s$^{-1}$ at higher energies, representing the first COMPTEL constraints on soft $\gamma$-ray emission from this cluster. Similarly, the DAMPE likelihood analysis finds no significant GeV--TeV emission (broadband TS~$\sim 0$), with 95\% confidence level upper limits of $\sim 10^{-12}$-$10^{-10}$~erg~cm$^{-2}$~s$^{-1}$ across 3~GeV--1~TeV.

\rthis{In comparison, recent Fermi-LAT analyses report GeV signals with
marginal significance ($4$--$4.5\sigma$), having integrated photon
fluxes $\sim 10^{-9}~\rm{ph~cm^{-2}~s^{-1}}$
\citep{Harale2025,Li2026}. The latter work also identified several
PMN radio sources \citep{Wright1994} within the $95\%$ localization
region of the excess, including two sources located within $0.2^\circ$
of the reported $\gamma$-ray position. Radio-loud AGN and, in
particular, relativistic jets are well-established sources of GeV
$\gamma$-ray emission, making these nearby radio sources plausible
alternative point-source contributors to the reported excess. However,
their spatial proximity alone does not establish that they are
responsible for the emission. Our DAMPE upper limits in the
$\sim$3--10~GeV range do not independently rule out the reported
Fermi-LAT excess, but our analysis finds no statistically significant
GeV--TeV $\gamma$-ray emission from Abell~119.}

\rthis{To report these constraints in a holistic fashion, we constructed a combined SED of Abell~119 spanning over seven decades in energy, incorporating upper limits from Swift/BAT, INTEGRAL/ISGRI, COMPTEL, and DAMPE alongside the tentative Fermi-LAT measurement of \citet{Harale2025}. This representation enables a direct visual and quantitative comparison between the tentative GeV excess and the non-thermal constraints at lower and higher energies. The different levels of the upper limits primarily reflect the energy-dependent sensitivities of the individual instruments and do not, by themselves, indicate a preferred energy range for a putative non-thermal component. The combined constraints therefore provide broadband limits on any non-thermal emission from Abell~119, while the reported GeV excess remains unconfirmed by independent observations.}


\rthis{Finally, we note that no suitable archival NuSTAR observations covering Abell~119 were found.
Given the higher sensitivity of NuSTAR in the hard X-ray band from 3-79 keV compared to INTEGRAL/IBIS or Swift-BAT, future observations with NuSTAR could provide
complementary and potentially more stringent constraints on any
non-thermal hard X-ray emission from this cluster.} Definitive conclusions will require next-generation MeV observatories such as COSI, AMEGO-X,  as well as improved GeV--TeV facilities (such as CTA), which will help determine whether genuine non-thermal emission is present in Abell~119.

\section*{ACKNOWLEDGEMENTS}
SM thanks the Ministry of Education (MoE), Government of India, for financial support through a research fellowship that enabled this work. This research is based on observations with INTEGRAL, an ESA project with instruments and a Science Data Centre funded by ESA Member States (particularly the PI countries: Denmark, France, Germany, Italy, Switzerland, and Spain), as well as the Czech Republic and Poland, with the participation of Russia and the USA. We acknowledge the use of public data from the INTEGRAL Science Data Centre (ISDC). This work also uses archival COMPTEL data obtained from the Compton Gamma Ray Observatory (CGRO) mission archive. We also acknowledge the use of data from the Dark Matter Particle Explorer (DAMPE) mission, supported by the Strategic Priority Program on Space Science, and the data services provided by the National Space Science Data Center of China.  \rthis{Finally, we acknowledge the anonymous referee for constructive feedback on the manuscript.}

\bibliographystyle{mn2e}
\bibliography{references}

@ARTICLE{Ackermann2014,
       author = {{Ackermann}, M. and {Ajello}, M. and {Albert}, A. and {Allafort}, A. and {Atwood}, W.~B. and {Baldini}, L. and {Ballet}, J. and {Barbiellini}, G. and {Bastieri}, D. and {Bechtol}, K. and {Bellazzini}, R. and {Bloom}, E.~D. and {Bonamente}, E. and {Bottacini}, E. and {Brandt}, T.~J. and {Bregeon}, J. and {Brigida}, M. and {Bruel}, P. and {Buehler}, R. and {Buson}, S. and {Caliandro}, G.~A. and {Cameron}, R.~A. and {Caraveo}, P.~A. and {Cavazzuti}, E. and {Chaves}, R.~C.~G. and {Chiang}, J. and {Chiaro}, G. and {Ciprini}, S. and {Claus}, R. and {Cohen-Tanugi}, J. and {Conrad}, J. and {D'Ammando}, F. and {de Angelis}, A. and {de Palma}, F. and {Dermer}, C.~D. and {Digel}, S.~W. and {Drell}, P.~S. and {Drlica-Wagner}, A. and {Favuzzi}, C. and {Franckowiak}, A. and {Funk}, S. and {Fusco}, P. and {Gargano}, F. and {Gasparrini}, D. and {Germani}, S. and {Giglietto}, N. and {Giordano}, F. and {Giroletti}, M. and {Godfrey}, G. and {Gomez-Vargas}, G.~A. and {Grenier}, I.~A. and {Guiriec}, S. and {Gustafsson}, M. and {Hadasch}, D. and {Hayashida}, M. and {Hewitt}, J. and {Hughes}, R.~E. and {Jeltema}, T.~E. and {J{\'o}hannesson}, G. and {Johnson}, A.~S. and {Kamae}, T. and {Kataoka}, J. and {Kn{\"o}dlseder}, J. and {Kuss}, M. and {Lande}, J. and {Larsson}, S. and {Latronico}, L. and {Llena Garde}, M. and {Longo}, F. and {Loparco}, F. and {Lovellette}, M.~N. and {Lubrano}, P. and {Mayer}, M. and {Mazziotta}, M.~N. and {McEnery}, J.~E. and {Michelson}, P.~F. and {Mitthumsiri}, W. and {Mizuno}, T. and {Monzani}, M.~E. and {Morselli}, A. and {Moskalenko}, I.~V. and {Murgia}, S. and {Nemmen}, R. and {Nuss}, E. and {Ohsugi}, T. and {Orienti}, M. and {Orlando}, E. and {Ormes}, J.~F. and {Perkins}, J.~S. and {Pesce-Rollins}, M. and {Piron}, F. and {Pivato}, G. and {Rain{\`o}}, S. and {Rando}, R. and {Razzano}, M. and {Razzaque}, S. and {Reimer}, A. and {Reimer}, O. and {Ruan}, J. and {S{\'a}nchez-Conde}, M. and {Schulz}, A. and {Sgr{\`o}}, C. and {Siskind}, E.~J. and {Spandre}, G. and {Spinelli}, P. and {Storm}, E. and {Strong}, A.~W. and {Suson}, D.~J. and {Takahashi}, H. and {Thayer}, J.~G. and {Thayer}, J.~B. and {Thompson}, D.~J. and {Tibaldo}, L. and {Tinivella}, M. and {Torres}, D.~F. and {Troja}, E. and {Uchiyama}, Y. and {Usher}, T.~L. and {Vandenbroucke}, J. and {Vianello}, G. and {Vitale}, V. and {Winer}, B.~L. and {Wood}, K.~S. and {Zimmer}, S. and {Fermi-LAT Collaboration} and {Pinzke}, A. and {Pfrommer}, C.},
        title = "{Search for Cosmic-Ray-induced Gamma-Ray Emission in Galaxy Clusters}",
      journal = {\apj},
         year = 2014,
        month = may,
       volume = {787},
       number = {1},
          eid = {18},
        pages = {18},
          doi = {10.1088/0004-637X/787/1/18},
archivePrefix = {arXiv},
       eprint = {1308.5654},
 primaryClass = {astro-ph.HE},
       adsurl = {https://ui.adsabs.harvard.edu/abs/2014ApJ...787...18A}
}

@ARTICLE{Chang2017,
       author = {{Chang}, J. and {Ambrosi}, G. and {An}, Q. and {Asfandiyarov}, R. and {Azzarello}, P. and {Bernardini}, P. and {Bertucci}, B. and {Cai}, M.~S. and {Caragiulo}, M. and {Chen}, D.~Y. and {Chen}, H.~F. and {Chen}, J.~L. and {Chen}, W. and {Cui}, M.~Y. and {Cui}, T.~S. and {D'Amone}, A. and {De Benedittis}, A. and {De Mitri}, I. and {Di Santo}, M. and {Dong}, J.~N. and {Dong}, T.~K. and {Dong}, Y.~F. and {Dong}, Z.~X. and {Donvito}, G. and {Droz}, D. and {Duan}, K.~K. and {Duan}, J.~L. and {Duranti}, M. and {D'Urso}, D. and {Fan}, R.~R. and {Fan}, Y.~Z. and {Fang}, F. and {Feng}, C.~Q. and {Feng}, L. and {Fusco}, P. and {Gallo}, V. and {Gan}, F.~J. and {Gan}, W.~Q. and {Gao}, M. and {Gao}, S.~S. and {Gargano}, F. and {Gong}, K. and {Gong}, Y.~Z. and {Guo}, J.~H. and {Hu}, Y.~M. and {Huang}, G.~S. and {Huang}, Y.~Y. and {Ionica}, M. and {Jiang}, D. and {Jiang}, W. and {Jin}, X. and {Kong}, J. and {Lei}, S.~J. and {Li}, S. and {Li}, X. and {Li}, W.~L. and {Li}, Y. and {Liang}, Y.~F. and {Liang}, Y.~M. and {Liao}, N.~H. and {Liu}, Q.~Z. and {Liu}, H. and {Liu}, J. and {Liu}, S.~B. and {Liu}, Q.~Z. and {Liu}, W.~Q. and {Liu}, Y. and {Loparco}, F. and {L{\"u}}, J. and {Ma}, M. and {Ma}, P.~X. and {Ma}, S.~Y. and {Ma}, T. and {Ma}, X.~Q. and {Ma}, X.~Y. and {Marsella}, G. and {Mazziotta}, M.~N. and {Mo}, D. and {Miao}, T.~T. and {Niu}, X.~Y. and {Pohl}, M. and {Peng}, X.~Y. and {Peng}, W.~X. and {Qiao}, R. and {Rao}, J.~N. and {Salinas}, M.~M. and {Shang}, G.~Z. and {Shen}, W.~H. and {Shen}, Z.~Q. and {Shen}, Z.~T. and {Song}, J.~X. and {Su}, H. and {Su}, M. and {Sun}, Z.~Y. and {Surdo}, A. and {Teng}, X.~J. and {Tian}, X.~B. and {Tykhonov}, A. and {Vagelli}, V. and {Vitillo}, S. and {Wang}, C. and {Wang}, Chi and {Wang}, H. and {Wang}, H.~Y. and {Wang}, J.~Z. and {Wang}, L.~G. and {Wang}, Q. and {Wang}, S. and {Wang}, X.~H. and {Wang}, X.~L. and {Wang}, Y.~F. and {Wang}, Y.~P. and {Wang}, Y.~Z. and {Wen}, S.~C. and {Wang}, Z.~M. and {Wei}, D.~M. and {Wei}, J.~J. and {Wei}, Y.~F. and {Wu}, D. and {Wu}, J. and {Wu}, S.~S. and {Wu}, X. and {Xi}, K. and {Xia}, Z.~Q. and {Xin}, Y.~L. and {Xu}, H.~T. and {Xu}, Z.~L. and {Xu}, Z.~Z. and {Xue}, G.~F. and {Yang}, H.~B. and {Yang}, J. and {Yang}, P. and {Yang}, Y.~Q. and {Yang}, Z.~L. and {Yao}, H.~J. and {Yu}, Y.~H. and {Yuan}, Q. and {Yue}, C. and {Zang}, J.~J. and {Zhang}, C. and {Zhang}, D.~L. and {Zhang}, F. and {Zhang}, J.~B. and {Zhang}, J.~Y. and {Zhang}, J.~Z. and {Zhang}, L. and {Zhang}, P.~F. and {Zhang}, S.~X. and {Zhang}, W.~Z. and {Zhang}, Y. and {Zhang}, Y.~J. and {Zhang}, Y.~Q. and {Zhang}, Y.~L. and {Zhang}, Y.~P. and {Zhang}, Z. and {Zhang}, Z.~Y. and {Zhao}, H. and {Zhao}, H.~Y. and {Zhao}, X.~F. and {Zhou}, C.~Y. and {Zhou}, Y. and {Zhu}, X. and {Zhu}, Y. and {Zimmer}, S.},
        title = "{The DArk Matter Particle Explorer mission}",
      journal = {Astroparticle Physics},
         year = 2017,
        month = oct,
       volume = {95},
        pages = {6-24},
          doi = {10.1016/j.astropartphys.2017.08.005},
archivePrefix = {arXiv},
       eprint = {1706.08453},
 primaryClass = {astro-ph.IM},
       adsurl = {https://ui.adsabs.harvard.edu/abs/2017APh....95....6C}
}

@ARTICLE{Li2026,
       author = {{Li}, Shang and {Han}, Feng},
        title = "{Search for {\ensuremath{\gamma}}-Ray Emission from Cluster of Galaxies with Fermi-LAT Data}",
      journal = {\apj},
         year = 2026,
        month = feb,
       volume = {997},
       number = {2},
          eid = {227},
        pages = {227},
          doi = {10.3847/1538-4357/ae2bda},
       adsurl = {https://ui.adsabs.harvard.edu/abs/2026ApJ...997..227L}
}

@ARTICLE{Ambrosi2019,
       author = {{Ambrosi}, G. and {An}, Q. and {Asfandiyarov}, R. and {Azzarello}, P. and {Bernardini}, P. and {Cai}, M.~S. and {Caragiulo}, M. and {Chang}, J. and {Chen}, D.~Y. and {Chen}, H.~F. and {Chen}, J.~L. and {Chen}, W. and {Cui}, M.~Y. and {Cui}, T.~S. and {Dai}, H.~T. and {D'Amone}, A. and {De Benedittis}, A. and {De Mitri}, I. and {Ding}, M. and {Di Santo}, M. and {Dong}, J.~N. and {Dong}, T.~K. and {Dong}, Y.~F. and {Dong}, Z.~X. and {Droz}, D. and {Duan}, K.~K. and {Duan}, J.~L. and {D'Urso}, D. and {Fan}, R.~R. and {Fan}, Y.~Z. and {Fang}, F. and {Feng}, C.~Q. and {Feng}, L. and {Fusco}, P. and {Gallo}, V. and {Gan}, F.~J. and {Gao}, M. and {Gao}, S.~S. and {Gargano}, F. and {Garrappa}, S. and {Gong}, K. and {Gong}, Y.~Z. and {Guo}, J.~H. and {Hu}, Y.~M. and {Huang}, G.~S. and {Huang}, Y.~Y. and {Ionica}, M. and {Jiang}, D. and {Jiang}, W. and {Jin}, X. and {Kong}, J. and {Lei}, S.~J. and {Li}, S. and {Li}, X. and {Li}, W.~L. and {Li}, Y. and {Liang}, Y.~F. and {Liang}, Y.~M. and {Liao}, N.~H. and {Liu}, C.~M. and {Liu}, H. and {Liu}, J. and {Liu}, S.~B. and {Liu}, W.~Q. and {Liu}, Y. and {Loparco}, F. and {Ma}, M. and {Ma}, P.~X. and {Ma}, S.~Y. and {Ma}, T. and {Ma}, X.~Q. and {Ma}, X.~Y. and {Marsella}, G. and {Mazziotta}, M.~N. and {Mo}, D. and {Niu}, X.~Y. and {Pan}, X. and {Peng}, X.~Y. and {Peng}, W.~X. and {Qiao}, R. and {Rao}, J.~N. and {Salinas}, M.~M. and {Shang}, G.~Z. and {Shen}, W.~H. and {Shen}, Z.~Q. and {Shen}, Z.~T. and {Song}, J.~X. and {Su}, H. and {Su}, M. and {Sun}, Z.~Y. and {Surdo}, A. and {Teng}, X.~J. and {Tian}, X.~B. and {Tykhonov}, A. and {Vitillo}, S. and {Wang}, C. and {Wang}, H. and {Wang}, H.~Y. and {Wang}, J.~Z. and {Wang}, L.~G. and {Wang}, Q. and {Wang}, S. and {Wang}, X.~H. and {Wang}, X.~L. and {Wang}, Y.~F. and {Wang}, Y.~P. and {Wang}, Y.~Z. and {Wang}, Z.~M. and {Wen}, S.~C. and {Wei}, D.~M. and {Wei}, J.~J. and {Wei}, Y.~F. and {Wu}, D. and {Wu}, J. and {Wu}, L.~B. and {Wu}, S.~S. and {Wu}, X. and {Xi}, K. and {Xia}, Z.~Q. and {Xin}, Y.~L. and {Xu}, H.~T. and {Xu}, Z.~H. and {Xu}, Z.~L. and {Xu}, Z.~Z. and {Xue}, G.~F. and {Yang}, H.~B. and {Yang}, P. and {Yang}, Y.~Q. and {Yang}, Z.~L. and {Yao}, H.~J. and {Yu}, Y.~H. and {Yuan}, Q. and {Yue}, C. and {Zang}, J.~J. and {Zhang}, D.~L. and {Zhang}, F. and {Zhang}, J.~B. and {Zhang}, J.~Y. and {Zhang}, J.~Z. and {Zhang}, L. and {Zhang}, P.~F. and {Zhang}, S.~X. and {Zhang}, W.~Z. and {Zhang}, Y. and {Zhang}, Y.~J. and {Zhang}, Y.~Q. and {Zhang}, Y.~L. and {Zhang}, Y.~P. and {Zhang}, Z. and {Zhang}, Z.~Y. and {Zhao}, H. and {Zhao}, H.~Y. and {Zhao}, X.~F. and {Zhou}, C.~Y. and {Zhou}, Y. and {Zhu}, X. and {Zhu}, Y. and {Zimmer}, S.},
        title = "{The on-orbit calibration of DArk Matter Particle Explorer}",
      journal = {Astroparticle Physics},
         year = 2019,
        month = mar,
       volume = {106},
        pages = {18-34},
          doi = {10.1016/j.astropartphys.2018.10.006},
archivePrefix = {arXiv},
       eprint = {1907.02173},
 primaryClass = {astro-ph.IM},
       adsurl = {https://ui.adsabs.harvard.edu/abs/2019APh...106...18A}
}

@ARTICLE{Duan,
       author = {{Duan}, Kai-Kai and {Jiang}, Wei and {Liang}, Yun-Feng and {Shen}, Zhao-Qiang and {Xu}, Zun-Lei and {Fan}, Yi-Zhong and {Gargano}, Fabio and {Garrappa}, Simone and {Guo}, Dong-Ya and {Lei}, Shi-Jun and {Li}, Xiang and {Mazziotta}, Mario Nicola and {Munoz Salinas}, Maria Fernanda and {Su}, Meng and {Vagelli}, Valerio and {Yuan}, Qiang and {Yue}, Chuan and {Zimmer}, Stephan},
        title = "{DmpIRFs and DmpST: DAMPE instrument response functions and science tools for gamma-ray data analysis}",
      journal = {Research in Astronomy and Astrophysics},
         year = 2019,
        month = sep,
       volume = {19},
       number = {9},
          eid = {132},
        pages = {132},
          doi = {10.1088/1674-4527/19/9/132},
archivePrefix = {arXiv},
       eprint = {1904.13098},
 primaryClass = {astro-ph.HE},
       adsurl = {https://ui.adsabs.harvard.edu/abs/2019RAA....19..132D}
}

@ARTICLE{Chang2014,
       author = {{Chang}, Jin},
        title = "{Dark Matter Particle Explorer：The First Chinese Cosmic Ray and Hard {\ensuremath{\gamma}}-ray Detector in Space}",
      journal = {Chinese Journal of Space Science},
         year = 2014,
        month = jan,
       volume = {34},
       number = {5},
        pages = {550},
          doi = {10.11728/cjss2014.05.550},
       adsurl = {https://ui.adsabs.harvard.edu/abs/2014ChJSS..34..550C}
}

@ARTICLE{Mattox96,
       author = {{Mattox}, J.~R. and {Bertsch}, D.~L. and {Chiang}, J. and {Dingus}, B.~L. and {Digel}, S.~W. and {Esposito}, J.~A. and {Fierro}, J.~M. and {Hartman}, R.~C. and {Hunter}, S.~D. and {Kanbach}, G. and {Kniffen}, D.~A. and {Lin}, Y.~C. and {Macomb}, D.~J. and {Mayer-Hasselwander}, H.~A. and {Michelson}, P.~F. and {von Montigny}, C. and {Mukherjee}, R. and {Nolan}, P.~L. and {Ramanamurthy}, P.~V. and {Schneid}, E. and {Sreekumar}, P. and {Thompson}, D.~J. and {Willis}, T.~D.},
        title = "{The Likelihood Analysis of EGRET Data}",
      journal = {\apj},
         year = 1996,
        month = apr,
       volume = {461},
        pages = {396},
          doi = {10.1086/177068},
       adsurl = {https://ui.adsabs.harvard.edu/abs/1996ApJ...461..396M}
}

@ARTICLE{White1978,
       author = {{White}, S.~D.~M. and {Rees}, M.~J.},
        title = "{Core condensation in heavy halos: a two-stage theory for galaxy formation and clustering.}",
      journal = {\mnras},
         year = 1978,
        month = may,
       volume = {183},
        pages = {341-358},
          doi = {10.1093/mnras/183.3.341},
       adsurl = {https://ui.adsabs.harvard.edu/abs/1978MNRAS.183..341W}
}

@ARTICLE{Kravtsov2012,
       author = {{Kravtsov}, Andrey V. and {Borgani}, Stefano},
        title = "{Formation of Galaxy Clusters}",
      journal = {\araa},
         year = 2012,
        month = sep,
       volume = {50},
        pages = {353-409},
          doi = {10.1146/annurev-astro-081811-125502},
archivePrefix = {arXiv},
       eprint = {1205.5556},
 primaryClass = {astro-ph.CO},
       adsurl = {https://ui.adsabs.harvard.edu/abs/2012ARA&A..50..353K}
}

@ARTICLE{Allen2011,
       author = {{Allen}, Steven W. and {Evrard}, August E. and {Mantz}, Adam B.},
        title = "{Cosmological Parameters from Observations of Galaxy Clusters}",
      journal = {\araa},
         year = 2011,
        month = sep,
       volume = {49},
       number = {1},
        pages = {409-470},
          doi = {10.1146/annurev-astro-081710-102514},
archivePrefix = {arXiv},
       eprint = {1103.4829},
 primaryClass = {astro-ph.CO},
       adsurl = {https://ui.adsabs.harvard.edu/abs/2011ARA&A..49..409A}
}

@article{DAMPE,
    author = "Ambrosi, G. and others",
    collaboration = "DAMPE",
    title = "{Direct detection of a break in the teraelectronvolt cosmic-ray spectrum of electrons and positrons}",
    eprint = "1711.10981",
    archivePrefix = "arXiv",
    primaryClass = "astro-ph.HE",
    doi = "10.1038/nature24475",
    journal = "Nature",
    volume = "552",
    pages = "63--66",
    year = "2017"
}

@ARTICLE{Bocquet19,
       author = {{Bocquet}, S. and {Dietrich}, J.~P. and {Schrabback}, T. and {Bleem}, L.~E. and {Klein}, M. and {Allen}, S.~W. and {Applegate}, D.~E. and {Ashby}, M.~L.~N. and {Bautz}, M. and {Bayliss}, M. and others},
        title = "{Cluster Cosmology Constraints from the 2500 deg$^{2}$ SPT-SZ Survey: Inclusion of Weak Gravitational Lensing Data from Magellan and the Hubble Space Telescope}",
      journal = {\apj},
         year = 2019,
        month = jun,
       volume = {878},
       number = {1},
          eid = {55},
        pages = {55},
          doi = {10.3847/1538-4357/ab1f10},
archivePrefix = {arXiv},
       eprint = {1812.01679},
 primaryClass = {astro-ph.CO},
       adsurl = {https://ui.adsabs.harvard.edu/abs/2019ApJ...878...55B}
}

@ARTICLE{Bleem15,
       author = {{Bleem}, L.~E. and {Stalder}, B. and {de Haan}, T. and {Aird}, K.~A. and
         {Allen}, S.~W. and {Applegate}, D.~E. and {Ashby}, M.~L.~N. and
         {Bautz}, M. and {Bayliss}, M. and {Benson}, B.~A. and {Bocquet}, S. and
         {Brodwin}, M. and {Carlstrom}, J.~E. and {Chang}, C.~L. and {Chiu}, I. and
         {Cho}, H.~M. and {Clocchiatti}, A. and {Crawford}, T.~M. and
         {Crites}, A.~T. and {Desai}, S. and {Dietrich}, J.~P. and
         {Dobbs}, M.~A. and {Foley}, R.~J. and {Forman}, W.~R. and
         {George}, E.~M. and {Gladders}, M.~D. and {Gonzalez}, A.~H. and
         {Halverson}, N.~W. and {Hennig}, C. and {Hoekstra}, H. and
         {Holder}, G.~P. and {Holzapfel}, W.~L. and {Hrubes}, J.~D. and
         {Jones}, C. and {Keisler}, R. and {Knox}, L. and {Lee}, A.~T. and
         {Leitch}, E.~M. and {Liu}, J. and {Lueker}, M. and {Luong-Van}, D. and
         {Mantz}, A. and {Marrone}, D.~P. and {McDonald}, M. and
         {McMahon}, J.~J. and {Meyer}, S.~S. and {Mocanu}, L. and {Mohr}, J.~J. and
         {Murray}, S.~S. and {Padin}, S. and {Pryke}, C. and {Reichardt}, C.~L. and
         {Rest}, A. and {Ruel}, J. and {Ruhl}, J.~E. and {Saliwanchik}, B.~R. and
         {Saro}, A. and {Sayre}, J.~T. and {Schaffer}, K.~K. and
         {Schrabback}, T. and {Shirokoff}, E. and {Song}, J. and
         {Spieler}, H.~G. and {Stanford}, S.~A. and {Staniszewski}, Z. and
         {Stark}, A.~A. and {Story}, K.~T. and {Stubbs}, C.~W. and {Vand
        erlinde}, K. and {Vieira}, J.~D. and {Vikhlinin}, A. and
         {Williamson}, R. and {Zahn}, O. and {Zenteno}, A.},
        title = "{Galaxy Clusters Discovered via the Sunyaev-Zel'dovich Effect in the 2500-Square-Degree SPT-SZ Survey}",
      journal = {\apjs},
         year = 2015,
        month = feb,
       volume = {216},
       number = {2},
          eid = {27},
        pages = {27},
          doi = {10.1088/0067-0049/216/2/27},
archivePrefix = {arXiv},
       eprint = {1409.0850},
 primaryClass = {astro-ph.CO},
       adsurl = {https://ui.adsabs.harvard.edu/abs/2015ApJS..216...27B}
}

@ARTICLE{Vikhlinin2014,
       author = {{Vikhlinin}, A.~A. and {Kravtsov}, A.~V. and {Markevich}, M.~L. and {Sunyaev}, R.~A. and {Churazov}, E.~M.},
        title = "{Clusters of galaxies}",
      journal = {Physics Uspekhi},
         year = 2014,
        month = apr,
       volume = {57},
       number = {4},
          eid = {317-341},
        pages = {317-341},
          doi = {10.3367/UFNe.0184.201404a.0339},
       adsurl = {https://ui.adsabs.harvard.edu/abs/2014PhyU...57..317V}
}

@ARTICLE{Desai2018,
       author = {{Desai}, Shantanu},
        title = "{Limit on graviton mass from galaxy cluster Abell 1689}",
      journal = {Physics Letters B},
         year = 2018,
        month = feb,
       volume = {778},
        pages = {325-331},
          doi = {10.1016/j.physletb.2018.01.052},
archivePrefix = {arXiv},
       eprint = {1708.06502},
 primaryClass = {astro-ph.CO},
       adsurl = {https://ui.adsabs.harvard.edu/abs/2018PhLB..778..325D}
}

@ARTICLE{Bora2021,
       author = {{Bora}, Kamal and {Desai}, Shantanu},
        title = "{Constraints on the variation of fine structure constant from joint SPT-SZ and XMM-Newton observations}",
      journal = {\jcap},
         year = 2021,
        month = feb,
       volume = {2021},
       number = {2},
          eid = {012},
        pages = {012},
          doi = {10.1088/1475-7516/2021/02/012},
archivePrefix = {arXiv},
       eprint = {2008.10541},
 primaryClass = {astro-ph.CO},
       adsurl = {https://ui.adsabs.harvard.edu/abs/2021JCAP...02..012B}
}

@ARTICLE{Bora2021b,
       author = {{Bora}, Kamal and {Desai}, Shantanu},
        title = "{A test of cosmic distance duality relation using SPT-SZ galaxy clusters, Type Ia supernovae, and cosmic chronometers}",
      journal = {\jcap},
         year = 2021,
        month = jun,
       volume = {2021},
       number = {6},
          eid = {052},
        pages = {052},
          doi = {10.1088/1475-7516/2021/06/052},
archivePrefix = {arXiv},
       eprint = {2104.00974},
 primaryClass = {astro-ph.CO},
       adsurl = {https://ui.adsabs.harvard.edu/abs/2021JCAP...06..052B}
}

@ARTICLE{Bora2022,
       author = {{Bora}, Kamal and {Holanda}, R.~F.~L. and {Desai}, Shantanu and {Pereira}, S.~H.},
        title = "{A test of the standard dark matter density evolution law using galaxy clusters and cosmic chronometers}",
      journal = {European Physical Journal C},
         year = 2022,
        month = jan,
       volume = {82},
       number = {1},
          eid = {17},
        pages = {17},
          doi = {10.1140/epjc/s10052-022-09987-3},
archivePrefix = {arXiv},
       eprint = {2106.15805},
 primaryClass = {astro-ph.CO},
       adsurl = {https://ui.adsabs.harvard.edu/abs/2022EPJC...82...17B}
}

@ARTICLE{Manna2024,
       author = {{Manna}, Siddhant and {Desai}, Shantanu},
        title = "{Search for GeV gamma-ray emission from SPT-SZ selected galaxy clusters with 15 years of Fermi-LAT data}",
      journal = {\jcap},
         year = 2024,
        month = jan,
       volume = {2024},
       number = {1},
          eid = {017},
        pages = {017},
          doi = {10.1088/1475-7516/2024/01/017},
archivePrefix = {arXiv},
       eprint = {2310.07519},
 primaryClass = {astro-ph.HE},
       adsurl = {https://ui.adsabs.harvard.edu/abs/2024JCAP...01..017M}
}

@ARTICLE{Manna2024b,
       author = {{Manna}, Siddhant and {Desai}, Shantanu},
        title = "{A pilot search for MeV gamma-ray emission from five galaxy clusters using archival COMPTEL data}",
      journal = {\jcap},
         year = 2024,
        month = may,
       volume = {2024},
       number = {5},
          eid = {013},
        pages = {013},
          doi = {10.1088/1475-7516/2024/05/013},
archivePrefix = {arXiv},
       eprint = {2401.13240},
 primaryClass = {astro-ph.HE},
       adsurl = {https://ui.adsabs.harvard.edu/abs/2024JCAP...05..013M}
}

@ARTICLE{Murase2013,
       author = {{Murase}, Kohta and {Beacom}, John F.},
        title = "{Galaxy clusters as reservoirs of heavy dark matter and high-energy cosmic rays: constraints from neutrino observations}",
      journal = {\jcap},
         year = 2013,
        month = feb,
       volume = {2013},
       number = {2},
          eid = {028},
        pages = {028},
          doi = {10.1088/1475-7516/2013/02/028},
archivePrefix = {arXiv},
       eprint = {1209.0225},
 primaryClass = {astro-ph.HE},
       adsurl = {https://ui.adsabs.harvard.edu/abs/2013JCAP...02..028M}
}

@ARTICLE{Condorelli2023,
       author = {{Condorelli}, Antonio and {Biteau}, Jonathan and {Adam}, Remi},
        title = "{Impact of Galaxy Clusters on the Propagation of Ultrahigh-energy Cosmic Rays}",
      journal = {\apj},
         year = 2023,
        month = nov,
       volume = {957},
       number = {2},
          eid = {80},
        pages = {80},
          doi = {10.3847/1538-4357/acfeef},
archivePrefix = {arXiv},
       eprint = {2309.04380},
 primaryClass = {astro-ph.HE},
       adsurl = {https://ui.adsabs.harvard.edu/abs/2023ApJ...957...80C}
}

@ARTICLE{Baghmanyan2022,
       author = {{Baghmanyan}, Vardan and {Zargaryan}, Davit and {Aharonian}, Felix and {Yang}, Ruizhi and {Casanova}, Sabrina and {Mackey}, Jonathan},
        title = "{Detailed study of extended {\ensuremath{\gamma}}-ray morphology in the vicinity of the Coma cluster with Fermi Large Area Telescope}",
      journal = {\mnras},
         year = 2022,
        month = oct,
       volume = {516},
       number = {1},
        pages = {562-571},
          doi = {10.1093/mnras/stac2266},
archivePrefix = {arXiv},
       eprint = {2110.00309},
 primaryClass = {astro-ph.HE},
       adsurl = {https://ui.adsabs.harvard.edu/abs/2022MNRAS.516..562B}
}

@ARTICLE{Ubertini2003,
       author = {{Ubertini}, P. and {Lebrun}, F. and {Di Cocco}, G. and {Bazzano}, A. and {Bird}, A.~J. and {Broenstad}, K. and {Goldwurm}, A. and {La Rosa}, G. and {Labanti}, C. and {Laurent}, P. and {Mirabel}, I.~F. and {Quadrini}, E.~M. and {Ramsey}, B. and {Reglero}, V. and {Sabau}, L. and {Sacco}, B. and {Staubert}, R. and {Vigroux}, L. and {Weisskopf}, M.~C. and {Zdziarski}, A.~A.},
        title = "{IBIS: The Imager on-board INTEGRAL}",
      journal = {\aap},
         year = 2003,
        month = nov,
       volume = {411},
        pages = {L131-L139},
          doi = {10.1051/0004-6361:20031224},
       adsurl = {https://ui.adsabs.harvard.edu/abs/2003A&A...411L.131U}
}

@ARTICLE{Winkler2003,
       author = {{Winkler}, C. and {Courvoisier}, T.~J.-L. and {Di Cocco}, G. and {Gehrels}, N. and {Gim{\'e}nez}, A. and {Grebenev}, S. and {Hermsen}, W. and {Mas-Hesse}, J.~M. and {Lebrun}, F. and {Lund}, N. and {Palumbo}, G.~G.~C. and {Paul}, J. and {Roques}, J.-P. and {Schnopper}, H. and {Sch{\"o}nfelder}, V. and {Sunyaev}, R. and {Teegarden}, B. and {Ubertini}, P. and {Vedrenne}, G. and {Dean}, A.~J.},
        title = "{The INTEGRAL mission}",
      journal = {\aap},
         year = 2003,
        month = nov,
       volume = {411},
        pages = {L1-L6},
          doi = {10.1051/0004-6361:20031288},
       adsurl = {https://ui.adsabs.harvard.edu/abs/2003A&A...411L...1W}
}

@ARTICLE{Curvoisier2003,
       author = {{Courvoisier}, T.~J.-L. and {Walter}, R. and {Beckmann}, V. and {Dean}, A.~J. and {Dubath}, P. and {Hudec}, R. and {Kretschmar}, P. and {Mereghetti}, S. and {Montmerle}, T. and {Mowlavi}, N. and {Paltani}, S. and {Preite Martinez}, A. and {Produit}, N. and {Staubert}, R. and {Strong}, A.~W. and {Swings}, J.-P. and {Westergaard}, N.~J. and {White}, N. and {Winkler}, C. and {Zdziarski}, A.~A.},
        title = "{The INTEGRAL Science Data Centre (ISDC)}",
      journal = {\aap},
         year = 2003,
        month = nov,
       volume = {411},
        pages = {L53-L57},
          doi = {10.1051/0004-6361:20031172},
archivePrefix = {arXiv},
       eprint = {astro-ph/0308047},
 primaryClass = {astro-ph},
       adsurl = {https://ui.adsabs.harvard.edu/abs/2003A&A...411L..53C}
}

@ARTICLE{Keshet,
       author = {{Keshet}, Uri},
        title = "{Galaxy-cluster-stacked Fermi-LAT. Part II. Extended central hadronic signal}",
      journal = {\jcap},
         year = 2025,
        month = oct,
       volume = {2025},
       number = {10},
          eid = {016},
        pages = {016},
          doi = {10.1088/1475-7516/2025/10/016},
archivePrefix = {arXiv},
       eprint = {2502.19494},
 primaryClass = {astro-ph.HE},
       adsurl = {https://ui.adsabs.harvard.edu/abs/2025JCAP...10..016K}
}

@ARTICLE{Mannastacked,
       author = {{Manna}, Siddhant and {Desai}, Shantanu},
        title = "{A stacked analysis of GeV gamma-ray emission from SPT-SZ galaxy clusters with 16 years of Fermi-LAT data}",
      journal = {Physics of the Dark Universe},
         year = 2025,
        month = sep,
       volume = {49},
          eid = {101966},
        pages = {101966},
          doi = {10.1016/j.dark.2025.101966},
archivePrefix = {arXiv},
       eprint = {2502.15235},
 primaryClass = {astro-ph.HE},
       adsurl = {https://ui.adsabs.harvard.edu/abs/2025PDU....4901966M}
}

@ARTICLE{Judit,
       author = {{P{\'e}rez-Romero}, Judit and {di Mauro}, Mattia and {Adam}, R{\'e}mi and {S{\'a}nchez-Conde}, Miguel {\'A}. and {Zaharijas}, Gabrijela},
        title = "{Search for cosmic-ray induced gamma-ray emission from local galaxy clusters using Fermi-LAT data}",
      journal = {arXiv e-prints},
         year = 2025,
        month = sep,
          eid = {arXiv:2509.15720},
        pages = {arXiv:2509.15720},
          doi = {10.48550/arXiv.2509.15720},
archivePrefix = {arXiv},
       eprint = {2509.15720},
 primaryClass = {astro-ph.HE},
       adsurl = {https://ui.adsabs.harvard.edu/abs/2025arXiv250915720P}
}

@ARTICLE{Manna2025i,
       author = {{Manna}, Siddhant and {Desai}, Shantanu and {Krivonos}, Roman A.},
        title = "{A Search for Hard X-ray/Soft $γ$-ray Emission from SPT-CL J2012-5649 (Abell 3667) Using INTEGRAL/ISGRI}",
      journal = {arXiv e-prints},
         year = 2025,
        month = dec,
          eid = {arXiv:2512.12616},
        pages = {arXiv:2512.12616},
          doi = {10.48550/arXiv.2512.12616},
archivePrefix = {arXiv},
       eprint = {2512.12616},
 primaryClass = {astro-ph.HE},
       adsurl = {https://ui.adsabs.harvard.edu/abs/2025arXiv251212616M}
}

@ARTICLE{Lebrun2003,
       author = {{Lebrun}, F. and {Leray}, J.~P. and {Lavocat}, P. and {Cr{\'e}tolle}, J. and {Arqu{\`e}s}, M. and {Blondel}, C. and {Bonnin}, C. and {Bou{\`e}re}, A. and {Cara}, C. and {Chaleil}, T. and {Daly}, F. and {Desages}, F. and {Dzitko}, H. and {Horeau}, B. and {Laurent}, P. and {Limousin}, O. and {Mathy}, F. and {Mauguen}, V. and {Meignier}, F. and {Molini{\'e}}, F. and {Poindron}, E. and {Rouger}, M. and {Sauvageon}, A. and {Tourrette}, T.},
        title = "{ISGRI: The INTEGRAL Soft Gamma-Ray Imager}",
      journal = {\aap},
         year = 2003,
        month = nov,
       volume = {411},
        pages = {L141-L148},
          doi = {10.1051/0004-6361:20031367},
archivePrefix = {arXiv},
       eprint = {astro-ph/0310362},
 primaryClass = {astro-ph},
       adsurl = {https://ui.adsabs.harvard.edu/abs/2003A&A...411L.141L}
}

@ARTICLE{Bird2016,
       author = {{Bird}, A.~J. and {Bazzano}, A. and {Malizia}, A. and {Fiocchi}, M. and {Sguera}, V. and {Bassani}, L. and {Hill}, A.~B. and {Ubertini}, P. and {Winkler}, C.},
        title = "{The IBIS Soft Gamma-Ray Sky after 1000 Integral Orbits}",
      journal = {\apjs},
         year = 2016,
        month = mar,
       volume = {223},
       number = {1},
          eid = {15},
        pages = {15},
          doi = {10.3847/0067-0049/223/1/15},
archivePrefix = {arXiv},
       eprint = {1601.06074},
 primaryClass = {astro-ph.HE},
       adsurl = {https://ui.adsabs.harvard.edu/abs/2016ApJS..223...15B}
}

@ARTICLE{Krivonos2010,
       author = {{Krivonos}, R. and {Tsygankov}, S. and {Revnivtsev}, M. and {Grebenev}, S. and {Churazov}, E. and {Sunyaev}, R.},
        title = "{INTEGRAL/IBIS 7-year All-Sky Hard X-Ray Survey. II. Catalog of sources}",
      journal = {\aap},
         year = 2010,
        month = nov,
       volume = {523},
          eid = {A61},
        pages = {A61},
          doi = {10.1051/0004-6361/201014935},
archivePrefix = {arXiv},
       eprint = {1006.4437},
 primaryClass = {astro-ph.HE},
       adsurl = {https://ui.adsabs.harvard.edu/abs/2010A&A...523A..61K}
}

@article{Wilks1938,
    author = "Wilks, S. S.",
    title = "{The Large-Sample Distribution of the Likelihood Ratio for Testing Composite Hypotheses}",
    doi = "10.1214/aoms/1177732360",
    journal = "Annals Math. Statist.",
    volume = "9",
    number = "1",
    pages = "60--62",
    year = "1938"
}

@ARTICLE{Lyons2013,
       author = {{Lyons}, Louis},
        title = "{Discovering the Significance of 5 sigma}",
      journal = {arXiv e-prints},
         year = 2013,
        month = oct,
          eid = {arXiv:1310.1284},
        pages = {arXiv:1310.1284},
          doi = {10.48550/arXiv.1310.1284},
archivePrefix = {arXiv},
       eprint = {1310.1284},
 primaryClass = {physics.data-an},
       adsurl = {https://ui.adsabs.harvard.edu/abs/2013arXiv1310.1284L}
}

@ARTICLE{Schoenfelder1993,
       author = {{Schoenfelder}, V. and {Aarts}, H. and {Bennett}, K. and {de Boer}, H. and {Clear}, J. and {Collmar}, W. and {Connors}, A. and {Deerenberg}, A. and {Diehl}, R. and {von Dordrecht}, A. and {den Herder}, J.~W. and {Hermsen}, W. and {Kippen}, M. and {Kuiper}, L. and {Lichti}, G. and {Lockwood}, J. and {Macri}, J. and {McConnell}, M. and {Morris}, D. and {Much}, R. and {Ryan}, J. and {Simpson}, G. and {Snelling}, M. and {Stacy}, G. and {Steinle}, H. and {Strong}, A. and {Swanenburg}, B.~N. and {Taylor}, B. and {de Vries}, C. and {Winkler}, C.},
        title = "{Instrument Description and Performance of the Imaging Gamma-Ray Telescope COMPTEL aboard the Compton Gamma-Ray Observatory}",
      journal = {\apjs},
         year = 1993,
        month = jun,
       volume = {86},
        pages = {657},
          doi = {10.1086/191794},
       adsurl = {https://ui.adsabs.harvard.edu/abs/1993ApJS...86..657S}
}

@ARTICLE{Knodlseder2016,
       author = {{Kn{\"o}dlseder}, J. and {Mayer}, M. and {Deil}, C. and {Cayrou}, J. -B. and {Owen}, E. and {Kelley-Hoskins}, N. and {Lu}, C. -C. and {Buehler}, R. and {Forest}, F. and {Louge}, T. and {Siejkowski}, H. and {Kosack}, K. and {Gerard}, L. and {Schulz}, A. and {Martin}, P. and {Sanchez}, D. and {Ohm}, S. and {Hassan}, T. and {Brau-Nogu{\'e}}, S.},
        title = "{GammaLib and ctools. A software framework for the analysis of astronomical gamma-ray data}",
      journal = {\aap},
         year = 2016,
        month = aug,
       volume = {593},
          eid = {A1},
        pages = {A1},
          doi = {10.1051/0004-6361/201628822},
archivePrefix = {arXiv},
       eprint = {1606.00393},
 primaryClass = {astro-ph.IM},
       adsurl = {https://ui.adsabs.harvard.edu/abs/2016A&A...593A...1K}
}

@ARTICLE{Knodlseder2022,
       author = {{Kn{\"o}dlseder}, J. and {Collmar}, W. and {Jarry}, M. and {McConnell}, M.},
        title = "{COMPTEL data analysis using GammaLib and ctools}",
      journal = {\aap},
         year = 2022,
        month = sep,
       volume = {665},
          eid = {A84},
        pages = {A84},
          doi = {10.1051/0004-6361/202243826},
archivePrefix = {arXiv},
       eprint = {2207.13404},
 primaryClass = {astro-ph.IM},
       adsurl = {https://ui.adsabs.harvard.edu/abs/2022A&A...665A..84K}
}

@ARTICLE{Shrivastava2024,
       author = {{Shrivastava}, Niharika and {Manna}, Siddhant and {Desai}, Shantanu},
        title = "{Search for MeV gamma-ray emission from TeV bright red dwarfs with COMPTEL}",
      journal = {\jcap},
         year = 2024,
        month = sep,
       volume = {2024},
       number = {9},
          eid = {029},
        pages = {029},
          doi = {10.1088/1475-7516/2024/09/029},
archivePrefix = {arXiv},
       eprint = {2407.01060},
 primaryClass = {astro-ph.HE},
       adsurl = {https://ui.adsabs.harvard.edu/abs/2024JCAP...09..029S}
}

@ARTICLE{Manna2024c,
       author = {{Manna}, Siddhant and {Desai}, Shantanu},
        title = "{Search for GeV gamma-ray emission from SPT-CL J2012-5649 with six years of DAMPE data}",
      journal = {Journal of High Energy Astrophysics},
         year = 2024,
        month = nov,
       volume = {44},
        pages = {210-213},
          doi = {10.1016/j.jheap.2024.10.001},
archivePrefix = {arXiv},
       eprint = {2408.10983},
 primaryClass = {astro-ph.HE},
       adsurl = {https://ui.adsabs.harvard.edu/abs/2024JHEAp..44..210M}
}

@ARTICLE{Harale2025,
       author = {{Harale}, Gajanan D. and {Paul}, Surajit},
        title = "{Excess of diffuse gamma-ray emission detected from the galaxy cluster Abell 119 from 14-year Fermi-LAT data}",
      journal = {\prd},
         year = 2025,
        month = nov,
       volume = {112},
       number = {10},
          eid = {103025},
        pages = {103025},
          doi = {10.1103/gn1q-pzx3},
archivePrefix = {arXiv},
       eprint = {2511.15559},
 primaryClass = {astro-ph.HE},
       adsurl = {https://ui.adsabs.harvard.edu/abs/2025PhRvD.112j3025H}
}

@ARTICLE{Wright1994,
       author = {{Wright}, Alan E. and {Griffith}, Mark R. and {Burke}, B.~F. and {Ekers}, R.~D.},
        title = "{The Parkes-MIT-NRAO (PMN) Surveys. II. Source Catalog for the Southern Survey (-87 degrees -4pt.5 < delta < -37 degrees )}",
      journal = {\apjs},
         year = 1994,
        month = mar,
       volume = {91},
        pages = {111},
          doi = {10.1086/191939},
       adsurl = {https://ui.adsabs.harvard.edu/abs/1994ApJS...91..111W}
}

@ARTICLE{Wik2012,
       author = {{Wik}, Daniel R. and {Sarazin}, Craig L. and {Zhang}, Yu-Ying and {Baumgartner}, Wayne H. and {Mushotzky}, Richard F. and {Tueller}, Jack and {Okajima}, Takashi and {Clarke}, Tracy E.},
        title = "{The Swift Burst Alert Telescope Perspective on Non-thermal Emission in HIFLUGCS Galaxy Clusters}",
      journal = {\apj},
         year = 2012,
        month = mar,
       volume = {748},
       number = {1},
          eid = {67},
        pages = {67},
          doi = {10.1088/0004-637X/748/1/67},
archivePrefix = {arXiv},
       eprint = {1207.0506},
 primaryClass = {astro-ph.CO},
       adsurl = {https://ui.adsabs.harvard.edu/abs/2012ApJ...748...67W}
}

@ARTICLE{Smith2004,
       author = {{Smith}, Russell J. and {Hudson}, Michael J. and {Nelan}, Jenica E. and {Moore}, Stephen A.~W. and {Quinney}, Stephen J. and {Wegner}, Gary A. and {Lucey}, John R. and {Davies}, Roger L. and {Malecki}, Justin J. and {Schade}, David and {Suntzeff}, Nicholas B.},
        title = "{NOAO Fundamental Plane Survey. I. Survey Design, Redshifts, and Velocity Dispersion Data}",
      journal = {The Astronomical Journal},
         year = 2004,
        month = oct,
       volume = {128},
       number = {4},
        pages = {1558-1569},
          doi = {10.1086/423915},
       adsurl = {https://ui.adsabs.harvard.edu/abs/2004AJ....128.1558S}
}

@ARTICLE{Way1997,
       author = {{Way}, M.~J. and {Quintana}, H. and {Infante}, L.},
        title = "{The Dynamics of the cD Clusters Abell 119 and Abell 133}",
      journal = {arXiv e-prints},
         year = 1997,
        month = sep,
          eid = {astro-ph/9709036},
        pages = {astro-ph/9709036},
          doi = {10.48550/arXiv.astro-ph/9709036},
archivePrefix = {arXiv},
       eprint = {astro-ph/9709036},
 primaryClass = {astro-ph},
       adsurl = {https://ui.adsabs.harvard.edu/abs/1997astro.ph..9036W}
}

@ARTICLE{Lee2016,
       author = {{Lee}, Youngdae and {Rey}, Soo-Chang and {Hilker}, Michael and {Sheen}, Yun-Kyeong and {Yi}, Sukyoung K.},
        title = "{Galaxy Luminosity Function of the Dynamically Young Abell 119 Cluster: Probing the Cluster Assembly}",
      journal = {\apj},
         year = 2016,
        month = may,
       volume = {822},
       number = {2},
          eid = {92},
        pages = {92},
          doi = {10.3847/0004-637X/822/2/92},
archivePrefix = {arXiv},
       eprint = {1603.01556},
 primaryClass = {astro-ph.GA},
       adsurl = {https://ui.adsabs.harvard.edu/abs/2016ApJ...822...92L}
}

@ARTICLE{Markevitch1998,
       author = {{Markevitch}, Maxim and {Forman}, William R. and {Sarazin}, Craig L. and {Vikhlinin}, Alexey},
        title = "{The Temperature Structure of 30 Nearby Clusters Observed with ASCA: Similarity of Temperature Profiles}",
      journal = {\apj},
         year = 1998,
        month = aug,
       volume = {503},
       number = {1},
        pages = {77-96},
          doi = {10.1086/305976},
archivePrefix = {arXiv},
       eprint = {astro-ph/9711289},
 primaryClass = {astro-ph},
       adsurl = {https://ui.adsabs.harvard.edu/abs/1998ApJ...503...77M}
}

@ARTICLE{Watson2023,
       author = {{Watson}, Courtney B. and {Blanton}, Elizabeth L. and {Randall}, Scott W. and {Sarazin}, Craig L. and {Sarkar}, Arnab and {ZuHone}, John A. and {Douglass}, E.~M.},
        title = "{CHANDRA X-Ray Observations of A119: Cold Fronts and a Shock in an Evolved Off-axis Merger}",
      journal = {\apj},
         year = 2023,
        month = oct,
       volume = {955},
       number = {2},
          eid = {103},
        pages = {103},
          doi = {10.3847/1538-4357/acee74},
archivePrefix = {arXiv},
       eprint = {2308.04367},
 primaryClass = {astro-ph.GA},
       adsurl = {https://ui.adsabs.harvard.edu/abs/2023ApJ...955..103W}
}

@ARTICLE{Ghirlanda2011,
       author = {{Ghirlanda}, G. and {Ghisellini}, G. and {Tavecchio}, F. and {Foschini}, L. and {Bonnoli}, G.},
        title = "{The radio-{\ensuremath{\gamma}}-ray connection in Fermi blazars}",
      journal = {\mnras},
         year = 2011,
        month = may,
       volume = {413},
       number = {2},
        pages = {852-862},
          doi = {10.1111/j.1365-2966.2010.18173.x},
archivePrefix = {arXiv},
       eprint = {1007.2751},
 primaryClass = {astro-ph.HE},
       adsurl = {https://ui.adsabs.harvard.edu/abs/2011MNRAS.413..852G}
}

@ARTICLE{Krivonos2022,
       author = {{Krivonos}, Roman A. and {Sazonov}, Sergey Yu and {Kuznetsova}, Ekaterina A. and {Lutovinov}, Alexander A. and {Mereminskiy}, Ilya A. and {Tsygankov}, Sergey S.},
        title = "{INTEGRAL/IBIS 17-yr hard X-ray all-sky survey}",
      journal = {\mnras},
         year = 2022,
        month = mar,
       volume = {510},
       number = {4},
        pages = {4796-4807},
          doi = {10.1093/mnras/stab3751},
archivePrefix = {arXiv},
       eprint = {2111.02996},
 primaryClass = {astro-ph.HE},
       adsurl = {https://ui.adsabs.harvard.edu/abs/2022MNRAS.510.4796K}
}

@INPROCEEDINGS{Kirsch2005,
       author = {{Kirsch}, M.~G. and {Briel}, U.~G. and {Burrows}, D. and {Campana}, S. and {Cusumano}, G. and {Ebisawa}, K. and {Freyberg}, M.~J. and {Guainazzi}, M. and {Haberl}, F. and {Jahoda}, K. and {Kaastra}, J. and {Kretschmar}, P. and {Larsson}, S. and {Lubi{\'n}ski}, P. and {Mori}, K. and {Plucinsky}, P. and {Pollock}, A.~M. and {Rothschild}, R. and {Sembay}, S. and {Wilms}, J. and {Yamamoto}, M.},
        title = "{Crab: the standard x-ray candle with all (modern) x-ray satellites}",
    booktitle = {UV, X-Ray, and Gamma-Ray Space Instrumentation for Astronomy XIV},
         year = 2005,
       editor = {{Siegmund}, Oswald H.~W.},
       series = {Society of Photo-Optical Instrumentation Engineers (SPIE) Conference Series},
       volume = {5898},
        month = aug,
        pages = {22-33},
          doi = {10.1117/12.616893},
archivePrefix = {arXiv},
       eprint = {astro-ph/0508235},
 primaryClass = {astro-ph},
       adsurl = {https://ui.adsabs.harvard.edu/abs/2005SPIE.5898...22K}
}

\end{document}